\documentclass[11pt]{article}

\usepackage{amsmath}
\usepackage{amssymb}
\usepackage{lineno}
\usepackage{hyperref}

\usepackage{graphicx}

\usepackage{cite}
\usepackage{color} 

\usepackage[labelfont=bf,labelsep=period,justification=raggedright,font=small]{caption}

\makeatletter
\renewcommand{\@biblabel}[1]{\quad#1.}
\makeatother

\date{}

\makeatletter
\newcommand{\NRom}[1]{\uppercase\expandafter{\romannumeral #1\relax}}
\makeatother

\begin{document}
\begin{center}
{\Large
\textbf{Subcritical escape waves in schooling fish}
}
\ \vspace{10pt} \\

\noindent{Winnie Poel$^{a,b}$,
Bryan C. Daniels$^{c}$,
Matthew M. G. Sosna$^{d}$,
Colin R. Twomey$^{e}$,
Simon P. Leblanc$^{d,f}$,
Iain D. Couzin$^{g,h,j}$
Pawel Romanczuk$^{a,b,*}$
}
\vspace{8pt} \\
\footnotesize{
$^{a}$Institute for Theoretical Biology, Department of Biology, Humboldt Universit{\"a}t zu Berlin, D-10099 Berlin, Germany \\
$^{b}$Bernstein Center for Computational Neuroscience, Berlin, D-10115 Berlin, Germany \\
$^{c}$School for Complex Adaptive Systems, Arizona State University, Tempe, AZ 85287 \\ 
$^{d}$Department of Ecology and Evolutionary Biology, Princeton University, Princeton, NJ 08544 \\
$^{e}$Department of Biology, University of Pennsylvania, Philadelphia, PA, 19104 \\ 
$^{f}$Blend Labs, San Francisco, CA 94108\\
$^{g}$Department of Collective Behaviour, Max Planck Institute of Animal Behavior, D-78547 Konstanz, Germany \\
$^{h}$Department of Biology, University of Konstanz, D-78547 Konstanz, Germany \\
$^{j}$Centre for the Advanced Study of Collective Behaviour, University of Konstanz, D-78547 Konstanz, Germany \\
$\ast$ E-mail: pawel.romanczuk@hu-berlin.de}
\end{center}

\newcommand{\pr}[1]{\textcolor{red} {\textit{ [{PR: #1]}}}}
\newcommand{\ic}[1]{\textcolor[rgb]{0.1,0.45,0.85} {\textit{ [{IC: #1]}}}}
\newcommand{\mmgs}[1]{\textcolor[rgb]{0.8 0.4 0.0} {\textit{ [{MMGS: #1]}}}}

\newcommand{\crt}[1]{\textcolor[rgb]{.4 0.1 0.4} {\textit{ [{CRT: #1]}}}}
\newcommand{\spl}[1]{\textcolor[rgb]{0,0.45,0.1} {\textit{ [{SPL: #1]}}}}
\newcommand{\winnie}[1]{\textcolor [rgb]{0.02745098 0.61568627 0.72156863}{\textit {[{WP: #1]}}}}
\newcommand{\bcd}[1]{\textcolor[rgb]{0,0.75,0} {\textit {[{BCD: #1]}}}}
\newcommand{\checkthis}[1]{\textcolor{magenta}{[NEW] #1}}

\newcommand{\summary}[1]{\textcolor{blue}{{Summary: #1 }}}

\begin{abstract}
Living systems such as neuronal networks and animal groups process information about their environment via the dynamics of interacting units. These can transition between distinct macroscopic behaviors. Near such a transition (or critical point) collective computation is generally thought to be optimized, due to the associated maximal sensitivity to perturbations and fast dissemination of information. For biological systems, however, optimality depends on environmental context, making the flexible, context-dependent adoption of different distances to a critical point potentially more beneficial than its unique properties. 
Here, studying escape waves in schooling fish at two levels of perceived environmental risk, we investigate a) if and how distance to criticality is regulated in response to environmental changes and b) how the individual level benefits derived from special properties of the critical point compare to those achieved via regulation of the group’s distance to it.
 We find that the observed fish schools are subcritical (not maximally responsive and sensitive to environmental cues), but decrease their distance to criticality with increased perceived risk. Considering an individual’s hypothetical costs of two detection error types, we find that optimal distance to criticality depends on the riskiness and noisiness of the environment, which may explain the observed behavior. 
Our results highlight the benefit of evaluating biological consequences of different distances to criticality for individuals within animal collectives. This provides insights into the adaptive function of a collective system and motivates future questions about the evolutionary forces that brought the system to make this particular trade-off.

\end{abstract}
\section*{Introduction}
An important aspect of biological systems is their ability to process information about their environment in order to detect and appropriately react to changes within it. More specifically, in many such systems, such as gene regulatory networks \cite{Karlebach2008}, neuronal networks \cite{Bassett2006BrainNetworks,mora2011biological} or animal groups \cite{KrauseLivingInGroups,WardWebsterGroupLiving}, biological function relies on distributed processing of information through the collective dynamics of interacting, potentially heterogeneous, components or agents. 

In animal groups, individuals can benefit from social information provided by other group members \cite{LIMA199511}, but environmental and internal noise may also cause misleading social cues \cite{Rosenthal2015,SosTwoBak19}. Sharing the imperfect information of many agents can therefore not only increase each individual's likelihood to be informed about environmental changes (e.g. the presence of a predator) but also may risk an increase in false or irrelevant information being propagated, especially if behavioral decisions need to be fast \cite{Couzin2009CollectiveCognition}. Thus, collective biological systems need to develop mechanisms to navigate the trade-off between filtering out noise yet remaining sensitive to relevant information that may be accessible to only a few agents. 

This trade-off was suggested to be optimally managed in the vicinity of a so-called `critical point', \cite{mora2011biological, Munoz2018Crit} where statistical physics predicts the collective dynamics (of infinite systems) to be most sensitive to small differences in an external perturbation, potentially allowing them to react differently to noise and relevant environmental cues. Generally speaking, critical points (or manifolds) in the high-dimensional parameter space of a multi-agent system mark points where the system experiences a collective instability and undergoes a qualitative change in its aggregate dynamics, corresponding to a phase transition in an infinite system \cite{DanKraFla17,Munoz2018Crit,Gross2021ManyCriticalStates,goldenfeld1992}. For biological systems this change can have important functional and behavioral consequences, as in for instance the transition of the disordered movement of individuals into coordinated marching in locust nymphs at a critical density \cite{Buhl1402}. Similarly, transitions and criticality have been studied in various other biological contexts, ranging from neural activity and brain networks \cite{shew2011information,beggs2012being, Deco2012,priesemann2014spike,Gross2021ManyCriticalStates} to gene regulatory networks \cite{Daniels2018regnets} to collective behavior of cells \cite{Nadell2013cells}. In the context of animal collectives, signatures of near-criticality have been found in experiments in bird flocks \cite{bialek2014social}, mammals \cite{DanKraFla17}, insects \cite{attanasi2014finite,gelblum2015ant} and fish schools \cite{handegard2012}, while theoretical models have investigated maximal sensitivity of the system at the critical regime \cite{Calovi2015Pertubations, vanni2011criticality,hidalgo2014information,klamser2020collective}. The possible benefits of criticality to animals within collectives has predominantly been considered with respect to maximal sensitivity or flexibility of the system or the appearance of long-ranged correlations within it \cite{Ribeiro2021ScaleFree}. In this work, we directly combine data of an actual biological system with a modeling approach to investigate how being critical may allow a system to efficiently distinguish noise from relevant cues, such as associated with increased risk, and how distance to criticality may manage the trade-off associated to imperfect detection abilities. Furthermore, we analyze the effects that the resulting collective behavior has on individuals, recognizing that group-level optima suggested by criticality may not be evolutionarily stable with respect to individual-level adaptation \cite{klamser2020collective}.

Here, we investigate behavioral cascades that spread socially through animal groups. Specifically, we study escape waves in schools of juvenile fish (golden shiners, \textit{Notemigonus crysoleucas}) \cite{Rosenthal2015,SosTwoBak19,Herbert-Read15}, a system that allows us to explore core aspects of previous studies of criticality in biological systems, namely: 1) quantifying where in parameter space a particular system operates with respect to a critical transition, including identification of an aggregate variable (called an order parameter in physics) that is best suited to identify the transition in question (e.g. the average marching direction in case of the locust \cite{Buhl1402,Bazazi2011}); 2) identifying functional benefits of operating near criticality, e.g. in terms of collective computation and information processing \cite{kinouchi2006optimal,hidalgo2014information,klamser2020collective,Shew2013FunctionalBenefits} and 3) revealing the mechanisms that enable biological systems to control their critical behavior, and to adapt in order to function properly \cite{bornholdt2000topological,meisel2009adaptive,meisel2012failure,DanKraFla17}.

The escape waves studied here are an example of more general avalanche processes, which play an important role in the collective dynamics in many biological systems, including spike avalanches in neuronal networks and disease transmission in human or animal populations \cite{beggs2004neuronal,Hahn2010,beggs2012being,priesemann2014spike,KeehlingRohani2008,PasRomCas2015Epidemic}. In such processes a local change in the state of an individual unit (i.e. the spiking of a neuron or the infection of an individual) can trigger the same change in its neighbors, thus spreading through the system like an avalanche, with a rate of spreading that can decrease (subcritical) or increase (supercritical) as the avalanche grows. Across many living systems, we have evidence that the degree of such behavioral spreading is regulated. In neural cultures, adding biochemical regulators that modify excitation and inhibition can force the system to supercritical and subcritical states \cite{SheYanYu11}, while in macaques, key individuals have been shown to influence how conflict spreads through a colony \cite{DanKraFla17}. 

Juvenile golden shiners form coordinated schools in response to high mortality from predation risk in the wild and as an escape behavior show a startle response that spreads socially \cite{Domenici1165Shiners,Johannes89Shiners}. Previous findings have shown that the strength of the behavioral spreading in these schools is controlled by their spatial structure, predominantly defined via school density \cite{SosTwoBak19}, and encodes group members' perceived risk of the environment. Using a data-driven computational approach we extrapolate the existing experimental data from \cite{SosTwoBak19} to predict startle cascades across a wider range of school densities and investigate the hypothesis that there is a critical density exhibiting features of optimal information processing in the context of collective escape response in fish. We locate a maximum in collective sensitivity caused by a critical instability, yet the experimentally observed schools are not at this point but rather at different distances with respect to it depending on the perceived `riskiness' of the environment. Our results suggest that the costs of being critical, such as susceptibility to noise and associated energetic costs, may play an important role in animal groups and should be included alongside the benefits. Overall, our work highlights how regulating distance to criticality can manage the trade-off between robustness and sensitivity according to environmental context.

\section*{Results}
\subsection*{Modeling experimental cascade size distributions}
Our data come from two previous studies of startle cascades in quasi two-dimensional schools of juvenile fish (golden shiners, \textit{Notemigonus crysoleucas}) that were initiated by randomly occurring single startles \cite{Rosenthal2015, SosTwoBak19} (details in Methods section). The main text uses data from \cite{SosTwoBak19} of groups of $N=40$ fish under two experimental conditions that differ in group members' perceived risk of the environment, namely before and after an alarm substance was sprayed on the water surface. Following \cite{SosTwoBak19}, we refer to these conditions as \textit{`Baseline'} and \textit{`Alarmed'} (for larger groups, based on data from \cite{Rosenthal2015}, refer to section \ref{sec:rosenthal} of the SI). 
\begin{figure}[htpb]
\centering
\includegraphics[width=0.65\linewidth]{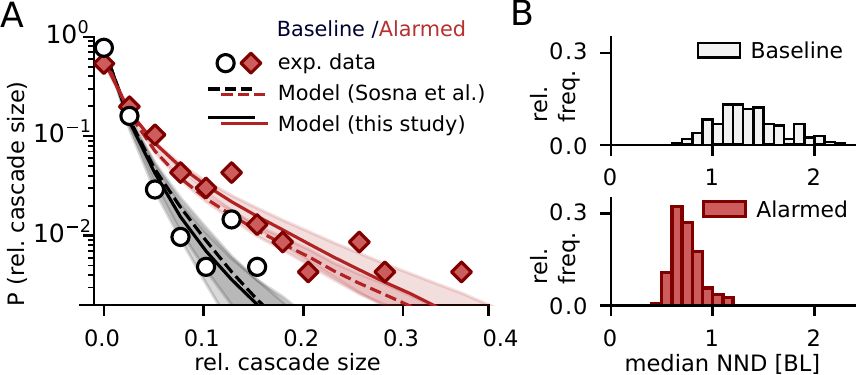}
\caption{Key aspects of `Baseline' and `Alarmed' experimental datasets from \cite{SosTwoBak19} differing in group members' perceived environmental risk: A) Observed cascade size distributions (data points). For increased perceived risk (`Alarmed', red diamonds) a larger average cascade size is observed. Both the model from \cite{SosTwoBak19} (dashed lines) and our approximation (solid lines) reproduce the observed cascade size distributions and agree within credible intervals (shaded areas). For more details on the model refer to the main text and Fig. \ref{fig:networks}. (B) Histograms characterizing school densities in the datasets via median nearest neighbor distance (NND). For the `Alarmed' condition (red, higher perceived risk) fish are closer to one another.}
\label{fig:expdata}
\end{figure}

Fig. \ref{fig:expdata} summarizes those experimental observations from \cite{SosTwoBak19} that are key to our study: on average schools in the `Alarmed' condition exhibit larger cascades (diamond and circles, Fig. \ref{fig:expdata}A) and a higher average spatial density (Fig. \ref{fig:expdata}B), characterized by median nearest neighbor distance (NND) measured in body length (BL), a measure which an individual may be able to perceive, and potentially control, through its social behavior. A high NND on average corresponds to low density and vice versa. The observed cascade size distributions were reproduced in \cite{Rosenthal2015, SosTwoBak19} by modeling a fractional contagion process spreading on statistically inferred interaction networks (Fig. \ref{fig:expdata}A, dashed lines). In this model, the nodes of the network represent individual fish and the static weighted and directed network links represent influence of one individual's startle behavior on that of another. 

A main finding of \cite{SosTwoBak19} is that the observed change in average cascade size cannot be explained by a change in individual responsiveness alone. Rather, change in the structure of the group is essential to explain the observed change in average cascade size. Next, we systematically study this structure-based control mechanism of spreading behavior and go beyond the limited experimentally observed samples of school densities shown in Fig. \ref{fig:expdata}B.

\subsection*{Predicting cascades beyond observed school densities}
The influence individual fish have on a neighbor's startle behavior is inferred statistically from experimental observations of individual's response rates to initial startles, with both \cite{SosTwoBak19,Rosenthal2015} finding measures based on inter-individual distances and visual properties most predictive of startle response. Here, we use these previously established predictive features to construct network links from the positions and orientations of individuals (see Methods, \eqref{eq:links} and SI). Through the underlying spatial positions of the individuals, networks are thus associated with a certain median NND and networks directly based on experimental data are thus limited to the two ranges of densities depicted in Fig. \ref{fig:expdata}B.

\begin{figure}[hbt]
    \centering
    \includegraphics[width=0.65\linewidth]{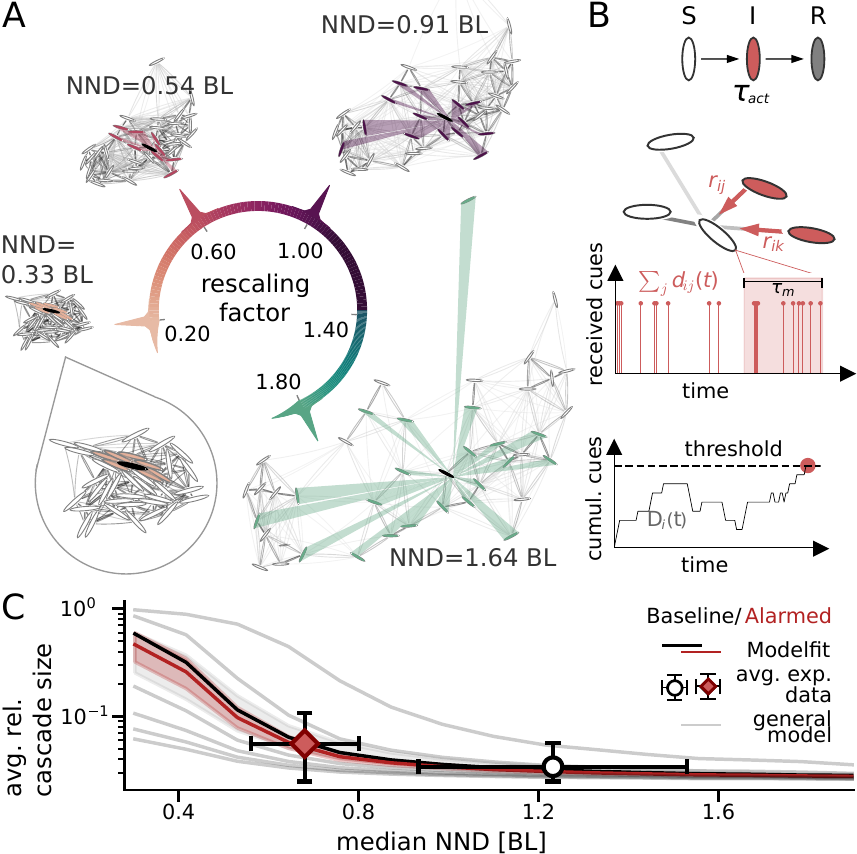}
    \caption{Predicting cascades across densities via a data driven computational approach: A) Examples of interaction networks obtained from rescaled position data for one experimental startle event with rescaling factors $d\in(0.2,0.6,1.0,1.8)$. The corresponding median nearest neighbor distance (NND) is noted next to each network. The darkness of lines between individuals represent link strength ($>0.01$ only, to keep the figure comprehensible). An example visual field (rays) originating from a focal individual (black) hitting visible neighbors (colored)) illustrates decrease in number of visible neighbors with increasing density due to occlusions. B) Scheme of SIR-type model dynamics: Observation of social cues over time triggers individual startling response mediated by strength of network links ($r_{ij}$) and individual responsiveness (threshold). C) Model predictions of avg. relative cascade size (lines) for different response thresholds show transition from local to global cascades with decreasing NND. Experimental observations from Fig. \ref{fig:expdata}B, shown here as averages $\pm$ one standard deviation (error bars) are best predicted by black and red curve respectively. Shaded areas are credible intervals of the model fit (grey: `Baseline', red: `Alarmed').}
    \label{fig:networks}
\end{figure}

To predict average cascade size beyond the experimentally observed school densities, we construct additional networks by rescaling positional data from experimental trials as illustrated in Fig. \ref{fig:networks}A. The rescaling changes the inter-individual distances and thus the associated density (noted next to each school), and also, correspondingly, visual interactions. We approximate the individuals as ellipses whose visual field we obtain using analytical calculations (see Methods and section \ref{sec:ellipse_validation} of the SI, where we also demonstrate the validity of this approximation). Overlaps of individuals are avoided via an automated routine that minimally adjusts rescaled positions and orientations. 
Fig. \ref{fig:networks}A illustrates the density dependent change of the visual field and the resulting interaction networks. A focal individual (black) has interaction links to all individuals which occupy an angular area above a threshold of 0.02 radians in the visual field (colored ellipses, threshold based on visual limits identified in \cite{Pita15} and as used in \cite{Rosenthal2015}; other visual thresholds do not affect the main results, see section \ref{sec:visuallimits} of the SI). Because visibility is not necessarily reciprocal, influence networks are directed and asymmetric. Link weight decreases with increasing distance between individuals (see Methods, eq. \eqref{eq:links}) and is indicated by link darkness, with darker lines corresponding to stronger links. Overall, with increasing density the interaction networks have fewer but stronger links (see SI for more details on network properties).

The spreading dynamics in our network model is equivalent to \cite{SosTwoBak19}, and the node behavior (based on a general model of behavioral contagion adopted from \cite{dodds2004universal,dodds2005generalized}) can be summarized as follows: An individual integrates stochastic activation cues that it receives from its startling neighbors over time and itself will startle when the amount of accumulated cues surpasses its response threshold (see Methods section for a detailed description of the SIR-type model and Fig. \ref{fig:networks}B for a model schematic.
As seen in Fig. \ref{fig:networks}C (grey lines), our model predicts that with increasing density the school's average relative cascade size (which we call its \textit{responsiveness}) in response to one initial startle changes from zero to one, with the exact position of the transition depending on the individual response threshold. In general, the collective dynamics thus can undergo a transition between a state of global cascades and one of no social spreading, mediated by the control parameter NND. We fit the threshold that best predicts the experimentally observed cascade sizes for each dataset via a maximum likelihood approach (see Fig. \ref{fig:ellipse_validation} for a plot of the log-likelihood and \ref{fig:expdata}A, solid lines, for the best model fit of the experimental data). The black and red curves in Fig. \ref{fig:networks}C show the prediction of the fitted models across a range of median NND.

\subsection*{Estimations of criticality indicate subcritical behavior}

\begin{figure}[htpb]
    \centering
    \includegraphics[width=0.65\linewidth]{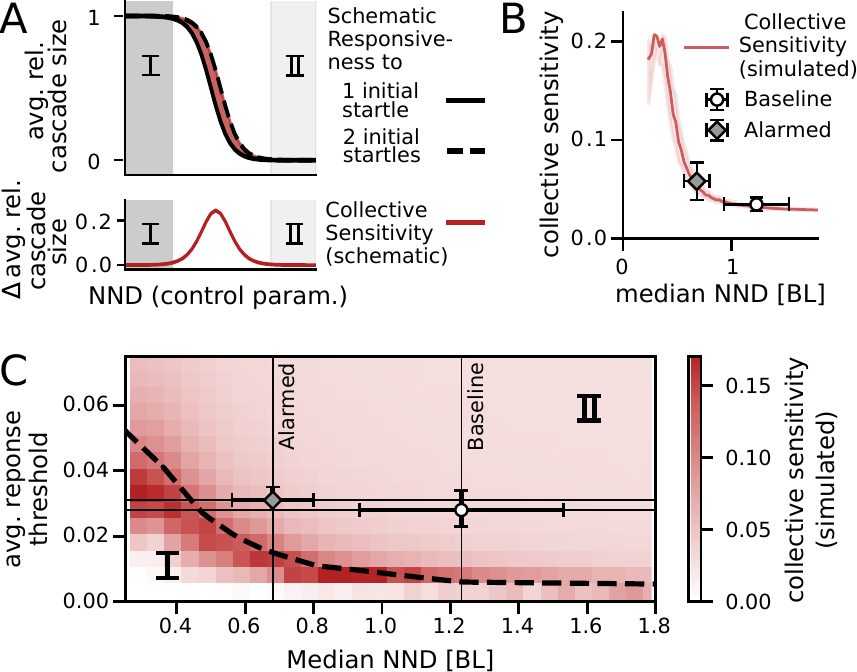}
    \caption{Estimating criticality via maximum sensitivity: A) Schematic sketch of transition to illustrate the concept behind B and C. Top: Responsiveness (avg. cascade size) to one and two initial startlers only differs around the transition (white background) which may enable different collective responses to noise and relevant cues. In both grey areas (\NRom{1} marking the supercritical and \NRom{2} the subcritical regime) any cue triggers the same response making distinction impossible. Bottom: collective sensitivity, defined as the difference between the responsiveness to two and one initial startles shows a peak at the transition.
    B) Collective sensitivity as a function of median nearest neighbor distance (NND). Lines: model predictions averaged over all networks. Shaded areas mark the credible interval of the model fit (see Fig. \ref{fig:ellipse_validation}). Lines end at low NND where physical bodies limit density. 
    The observed spreading is subcritical with the `Alarmed' condition closer to criticality than the `Baseline' condition as shown by markers with error bars representing simulation result averages over original scale networks (error bars indicate one standard deviation). C) Collective sensitivity as a function of median NND and average individual response threshold shows a peak close to the analytically estimated critical manifold ($b=1$, marked by dashed line), separating the subcritical (\NRom{1}) and supercritical (\NRom{2}) regime. Markers with error bars show averages over simulations on original scale networks and represent the observed schools' behavior.}
    \label{fig:sensitivity_1d}
\end{figure}

Fig. \ref{fig:sensitivity_1d}A shows a schematic visualization of the transition exhibited by our model which, for an infinitely large group, corresponds to a phase transition \cite{henkel2008non} -- a collective instability leading to a qualitative change in the aggregate dynamics. Far away from the transition \textit{any} perturbation (i.e. one or two initial startlers) will either cause (almost) the entire system to respond (\NRom{1}, supercritical regime, average relative cascade size is one for both the solid and dashed curve for the NND range with dark grey background) or will cause no response at all (\NRom{2}, subcritical regime, vanishing average relative cascade size, light grey background). Thus different perturbations cannot be distinguished by observing the collective response in these regimes. Around the critical point (white background) responses of all sizes are possible. Here, small differences in the perturbation, i.e. the number of initial startlers, lead to different average cascade sizes as qualitatively sketched in Fig. \ref{fig:sensitivity_1d}A for average cascades initiated by one ($c_1$) or two individuals ($c_2$). The difference between the two responses ($\Delta c = c_2-c_1$) peaks at the respective phase transition and is here referred to as the school's \textit{collective sensitivity}. In more general contexts, such increased sensitivity is thought to be beneficial for collective information processing. In our system increased sensitivity may allow the school to react differently to a predator cue than to noise cues and thus permit filtering, enabling large cascades to be triggered only by sufficiently strong cues.

Note that a phase transition becomes sharply well-defined only in an infinite system. In finite systems the transition between alternate global states becomes more gradual as system size decreases \cite{goldenfeld1992}.  Still, effects of the transition such as increased sensitivity persist near the infinite-size critical point.
Here, we will use the maximum of collective sensitivity as a functionally relevant indicator of near-criticality.  This is analogous in classical statistical mechanics to a maximum of susceptibility to a weak external perturbation acting as a characteristic feature of the quasi-critical point in finite systems \cite{henkel2008non}. 

We obtain the collective sensitivity via simulations of cascades initiated by different numbers of initial startles. While the experimental data only contain cascades initiated by one individual, using the model we can predict average cascade size triggered by any number of initial startles (see Methods). Specifically, we use the difference between 1 and 2 initial startlers to measure sensitivity to small perturbations (see Fig. \ref{fig:sensitivity_and_netneighbors} of the SI for other possible definitions of collective sensitivity, which show a similar qualitative behavior).
Assuming independent noise-induced spontaneous startles of individuals, the difference between a single startle and simultaneous startling of two fish will be most informative in identifying a potential threat in the case of a low signal to noise ratio, where only a few fish are likely to respond.

Fig. \ref{fig:sensitivity_1d}B depicts the dependence of the collective sensitivity on NND, obtained from averaging simulation results over all rescaled versions of experimentally observed networks. The line ends around a NND of 0.3 body lengths, where constructing 2d-networks of higher density is not possible (without assuming a highly-ordered, grid-like closest packing of ellipses) because of the physical bodies that are not allowed to overlap. We find that maximum sensitivity (of roughly 0.2 difference in relative cascade size between 1 and 2 initial startlers) occurs at a NND of about 0.3 to 0.4 BL, which coincides approximately with these highest simulated densities. 

The data points in Fig. \ref{fig:sensitivity_1d}B indicate averages over trials ($\pm$ one standard deviation) of the collective sensitivity obtained from simulations on the original scale networks. They thus characterize the behavior of the schools under the two different experimental conditions. The error bars of these data points are large due to the large range of observed school densities (refer to Fig. \ref{fig:densitymeasures} for a figure with the raw data). The two experimental data sets have an average collective sensitivity of  $0.035\pm0.007$ (`Baseline') and $0.058\pm0.019$ (`Alarmed') at average median NND of $1.23\pm0.3\text{\,BL}$ and $0.68\pm0.12\text{\,BL}$ respectively. Thus their sensitivity could be increased by a factor of 5.9 and 3.4 respectively by becoming critical. Instead, the observed schools are \textit{subcritical}, with NND large enough that cascades on average stay local. In addition, we observe that while schools do get closer to criticality when perceived risk is higher, the physical bodies act as a lower limit for NND, which prevents the school from becoming (super)critical by density modulation alone (this also holds for other measures of density, as shown in Fig. \ref{fig:densitymeasures}, and may extend to 3 dimensional schools, discussed in section \ref{sec:SI_3d} of the SI).

\subsection*{Regulation of distance to criticality}
Next, we explore whether an additional change of individual responsiveness (quantified by the average response threshold) can move the school to or across the critical manifold. Although \cite{SosTwoBak19} found no significant change in individual sensitivity in response to a higher perceived risk, these findings may not extend to multi-sensory cues or noisy environments \cite{munoz2012multisensory}. 
In Fig. \ref{fig:sensitivity_1d}C the collective sensitivity is plotted for varying both NND and the average response threshold and shows one clear band of maximum values.

To verify that the observed maximum in sensitivity is indeed caused by a critical instability, we derive a branching ratio measure as an analytical estimate of the critical point in an infinite system \cite{DanKraFla17}. The median branching ratio $b$ describes how small initial startling cascades, on average, tend to grow ($b>1$) or shrink ($b<1$), with $b=1$ marking the transition between these two distinct aggregate behaviors (and corresponding to a true phase transition in an infinite system; for details refer to methods). The black dashed line in Fig. \ref{fig:sensitivity_1d}C marks this analytically estimated critical manifold ($b=1$) separating the subcritical (\NRom{1}, $b<1$) and supercritical (\NRom{2}, $b>1$) regime (for the full dependency of the median branching ratio on NND and average response threshold refer to Fig. \ref{fig:branching} of the SI). The analytical estimate describes the observed maximum sensitivity well with remaining differences likely due to simplifying assumptions made in its derivation (see Methods section).

The black data points with error bars indicate the area in parameter space that best describes the two experimental data sets (horizontal bar: average density $\pm$ one standard deviation, vertical bar: optimal parameter fit and credible interval, see table \ref{tab:LL_ellipse_fovea}). Fig. \ref{fig:sensitivity_1d}B is thus a cross-section of this plot at the horizontal lines. Following the vertical lines in Fig. \ref{fig:sensitivity_1d}C shows that a decrease of average response threshold could bring the schools to criticality without a change in density, but the necessary threshold values lie outside of the credible interval. Overall, a change in density can move the school towards the transition and a change of individual responsiveness could even allow the school to cross to the supercritical regime (bottom left corner). Yet we find that the experimentally observed densities at both levels of perceived risk (and in the absence of a real predator) are \textit{not} located at a maximum of any sensitivity measure (Fig. \ref{fig:sensitivity_and_netneighbors}).

Having established that fish schools could potentially cross the transition (but remain subcritical in experiments), we next examine the potential benefits of criticality in this system from an information processing perspective and interpret them at the level of an individual fish.

\subsection*{Individual costs and benefits of criticality}
Golden shiner schools live in fission-fusion populations with fluctuating group membership \cite{Johannes89Shiners}. Given that fish do not consistently occupy certain positions in the school, we interpret the average relative cascade size as the probability that a fish will startle given some number of initial startlers. For example, if cascades initiated by one startle on average spread to 40\% of the school, we say that any fish has an average probability of 40\%  to respond to a cue that causes a single initial startle.

Fig. \ref{fig:sensitivity_1d}B can thus be interpreted as the difference in individual response probability to 1 and 2 initial startlers. It shows that, on average, the closer the school's NND is to the approximated critical point ($b=1$) the better the individual fish's response distinguishes numbers of initial startlers. The relevant distinction in this escape context is, however, whether the cascade was triggered by the detection of a predator or not, because the presence of a predator determines the optimal behavior for the individual. Ideally, all fish would escape when there is a predator (\textit{true positive}) and not do so when there is none (\textit{true negative}), but due to ambiguous cues (both environmental and social) this decision problem gives rise to two types of errors: A \textit{false positive} occurs when an individual startles in the absence of a predator and a \textit{false negative} when it fails to startle in the presence of a predator \cite{Wolf2013,Marshall2019}.

\begin{figure}[htpb]
    \centering
     \includegraphics[width=0.65\linewidth]{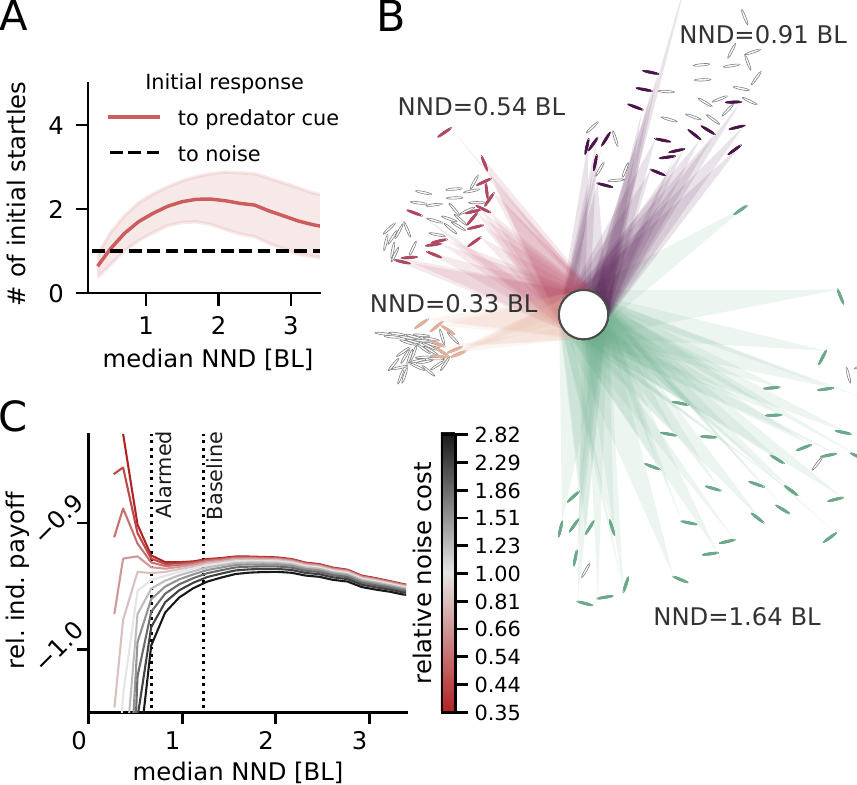}
    \caption{Hypothetical predator detection model reveals distance to criticality can manage trade-off between two types of errors: A) Initial response to predator and noise cue as function of school's median NND. Initial predator response is given by a fixed fraction of the average number of individuals that can see the predator. B) Visual predator detection for schools adjusted to have  different median NND (stated next to each school). Colored individuals are able to see the predator (white circle). Shaded areas illustrate the visual field similar to Fig. \ref{fig:networks}A. C) Relative payoff for an individual in a school in different environments (characterized by relative noise cost) as function of school's median NND. Averages of experimentally observed school densities are indicated (dotted vertical lines) as well as the estimated critical point (dashed vertical line). Depending on the environment different values of NND (i.e. different distances from criticality) maximize payoff. In risky environments (red), being highly responsive is more important than filtering, while in very noisy low-risk environments being critical is detrimental to the payoff (black curves). For intermediate values of relative noise cost (light grey/red) there are two maxima, one based on maximum sensitivity at criticality (left) and one based on maximum personal visual access to the predator (right).}
    \label{fig:fitness}
\end{figure}

We simulate the likelihood of the above errors for different distances to criticality via average cascade sizes triggered by different numbers of individuals which we associate to a noise cue or a predator cue. The simulated fraction of the group responding to a noise cue or \textit{not} responding to a predator cue gives the individual's probability of a false positive or a false negative respectively (see Fig. \ref{fig:errors}). Fig. \ref{fig:fitness}A shows the assumed number of initial startlers for each type of cue. To simulate false positives, cascades were initiated by one randomly chosen individual (dashed line). 
The number of initial startlers for a predator cue (solid line, Fig. \ref{fig:fitness}A) is obtained by modelling the school's ability to visually detect a predator as illustrated in Fig. \ref{fig:fitness}B for changing school density. The predator (white circle of diameter 3\,BL at equal distance to all school borders) can be seen by all colored individuals (ellipses). Shaded triangles illustrate the visual perception of the predator by the school (all rays emitted from an individual's eye that hit the predator).  Collective information processing is most beneficial for small signal to noise ratios \cite{daniels2019quantifying}, so we consider a cryptic predator, such that at any given time only a small fraction of individuals that can potentially see the predator identifies it as a threat and reacts with a startle. Here we use a ratio of $p_\text{detect}=0.1$ but the general results do not depend on the exact value (see Fig. \ref{fig:SI_visualfraction}). While occlusions limit detectability at low NND, the decrease of the curve in Fig. \ref{fig:fitness}A at high NND is due to an upper limit on the distance at which a predator is assumed to be visible. 

 To compare the benefits of different distances from criticality we introduce a \textit{relative individual payoff}. Simply put, this measure adds up the costs of the two different error types (loss of energy and time due to false positives or risk of death or injury due to false negatives) weighted by their likelihood of occurrence in a certain environment, which we obtain from simulations as described above. A single parameter, the \textit{relative noise cost}, captures both the riskiness and the noise level of the environment, with high values corresponding either to noisy or to safe environments and low values to environments with high predation or very low noise (for a detailed description of the relative payoff refer to methods). Here, noisy environments are characterized by low signal to noise ratio, where e.g. visual cues on the presence of a cryptic predator are difficult to distinguish from other environmental fluctuations.

Fig. \ref{fig:fitness}C shows the dependence of the relative payoff for an individual in a school of $N=40$ fish on NND for different relative noise cost. One observes three different kinds of dependency, outlined below.

1) For relatively safe and/or noisy environments (black curves) a single optimum appears at low density ($\text{NND}\approx 2\,\text{BL}$). The exact position of the maximum depends on the choice of predator distance and assumed maximal detection distance of an individual (see Fig. \ref{fig:vis_detection} and \ref{fig:SI_payoff_params} for a version with maxima at the observed averages of the experimental data set, marked by vertical dotted lines here). For these environments, getting closer to criticality decreases the payoff, which is dominated by the costs of false positives. The optimal spatial configuration (an intermediate NND) maximizes visual access to personal information about a potential predator while keeping social information relatively low and corresponds roughly to the maximum in Fig. \ref{fig:fitness}A. 

2) For environments with high predation and/or low noise (red curves), being closer to criticality increases the relative payoff, which is dominated by the costs of false negatives. In these environments, depending on $p_\text{detect}$, the maximal payoff lies either near the estimated critical point or past it in the inaccessible supercritical regime, where responsiveness to any cue is maximized (see figures \ref{fig:sensitivity_1d}A and \ref{fig:SI_payoff_lowthresh}). Under these conditions, individual response relies almost completely on social information. 

3) Only for intermediate relative noise cost (light gray to light red) does the close vicinity of the critical point yield the maximal payoff for the individual. Under these conditions, the individual can benefit from the behavioral contagion process' increased sensitivity to the number of initial startlers seen in Fig. \ref{fig:sensitivity_1d} (and \ref{fig:sensitivity_and_netneighbors} of the SI). In all other scenarios the benefit of an increased collective sensitivity near criticality is outweighed by either the increased false positive rate and decreased visual accessibility of the environment (compared to higher NND) or the decreased true positive rate (compared to lower NND). Only by including the sensory constraints on visual predator detection via the varying number of initial startles do we find the second maximum in the relative payoff in addition to the maximum near criticality. 

Changing the parameters of the visual predator detection can shift the visual-access-based maximum (see figures \ref{fig:SI_visualfraction} and \ref{fig:vis_detection}). We note that the average observed median NND for the `Baseline' and `Alarmed' datasets (indicated by dotted vertical lines) could be explained by an attempted optimization of individual payoff according to the perceived riskiness of the environment, with the `Baseline' condition corresponding to a higher and the `Alarmed' condition to a lower relative noise cost (see Fig. \ref{fig:SI_payoff_params}A). 

\section*{Discussion}
In this work we set out to (1) investigate criticality in the context of escape waves in animal groups and where the experimentally observed fish schools operate with respect to it (i.e. their distance to the critical point); (2) understand potential functional benefits of near-criticality in this context, especially with respect to the trade-off between robustness to noise versus sensitivity to environmental cues; (3) determine which individual or structural features control a group's distance to criticality. We combined experimental observations of alarms spreading through groups of juvenile golden shiners at two different average school densities with a computational model to predict the average collective response across different densities. This allowed us to identify a critical density at which a transition from predominantly local cascades to global cascades occurs. We find that under two experimental conditions the spreading of escape waves within the school is subcritical, with the schools in the `Alarmed' condition being closer to the transition, answering (1). 

Addressing (2), we then show that, as predicted by statistical mechanics, the critical manifold exhibits a high sensitivity to small differences in the strength of an initial perturbation (number of initial startlers). In order to understand biologically relevant functional benefits of near-criticality from the perspective of the individual fish, we then considered the costs associated with two possible errors in the binary escape decision, namely false positives and false negatives. Using simulated average cascade sizes and visual predator detection to infer individual's relative payoffs, we find that depending on the type of environment, being close to criticality may be beneficial or detrimental to individuals and that by changing their distance to criticality fish within a school can manage the trade-off between robustness and sensitivity according to their current environment. Focusing on visual predator detection we also identify a potential school density related trade-off between the acquisition of accurate personal and social information about the environment for the individual, where with increasing density the individual's private information about the environment decreases because of occlusions but the social information increases because of stronger interactions within the group. 

Regarding (3), we find that school density and individual responsiveness control the school's distance to criticality. Under the experimental conditions, the upper limit to school density due to the fishs' physical bodies prevents the schools from moving past the critical manifold to a supercritical state via modulation of density alone, potentially increasing robustness of contagion dynamics to density fluctuations. However, we can show that an increase in individual sensitivity would enable a transition to supercritical spreading.

The cascades we analyzed were triggered by random startles (false positives) in a laboratory environment with fish that were bred in captivity and in the absence of a real predator.
In insect swarms, it has been argued that large groups under natural conditions appear to operate near criticality \cite{attanasi2014finite}, yet small groups under laboratory conditions do not \cite{kelley2013emergent,puckett2014determining}. Whereas many questions remain open regarding the role of environmental factors, e.g. the role of wind gusts for the observed collective movement of such swarms under natural conditions, it is possibles that analogous effects could be observed in fish schools, with changes to the number of individuals or fluctuations in the natural environment driving the system closer to or potentially past the critical state. While we assume the initialization of a cascade by a certain number of fish at one point in time, the presence of an actual predator likely represents a temporally extended cue, which may cause additional (non-social) startles based on personal information and thus increase the average cascade size. Also, fish raised in non-laboratory conditions may develop different thresholds or interactions. Further investigations of startling cascades under ecologically relevant scenarios, including predation threat, environmental fluctuations, and a larger number of individuals will be key here. 
One can also imagine measuring cascade size distributions in the laboratory as a function of visual and/or acoustic noise.
 
 In our model we assume that the nature of interactions does not change with density. Acoustic and lateral-line sensory inputs could additionally influence interactions at high densities, potentially shifting the critical point. However, our previous work \cite{SosTwoBak19} showed that the functional form of the interactions does not change significantly for the naturally observed density adaptation in response to perceived risk. While our results do not preclude the possibility that other mechanisms may increase sensitivity further in high density situations, this would not change our main finding that the spreading is subcritical at the observed densities. 
 
 Previously an initial increase in density following an escape behavior has been identified as a key ingredient of escape waves in schooling fish \cite{Herbert-Read15}. Startles may systematically cause a change of local density and thus expedite or inhibit the spreading. Additionally, cascades may temporarily change the group's overall density and thus create a group level refractory period or an increased awareness to further signals. These aspects are not considered in our modelling approach which uses static interaction networks, which has been shown to be an appropriate approximation due to the high speed of spreading of cascades relative to changes in the interaction network topology \cite{Rosenthal2015}. Nevertheless, future studies should look at the full spatial dynamics starting with a comparison of the predicted interaction network at the beginning and the end of the cascade. If a systematic influence of behavioral spreading on the density can be found this might point to a regulatory mechanism of self-organized (sub)criticality that allows the school to control their interactions in order to remain responsive to relevant signals while filtering noise. 

While it was previously known that spatial structure of the school influences cascade size \cite{SosTwoBak19, Herbert-Read15}, this study is systematically exploring this influence, identify the critical density and place groups observed in experiments with respect to this transition. 
The investigation of two different environmental conditions and the corresponding observed change of distance to criticality sets this study apart from previous work on criticality in animal collectives \cite{bialek2014social,attanasi2014finite,DanKraFla17}. Our estimation of individual relative payoff offers a possible explanation of the observed subcritical spreading and emphasizes the importance of combining the collective description common to statistical physics with the individual perspective often taken in biological and psychological research on decision making. 

Our results suggest that the collective response of schooling fish minimizes the individuals' average cost due to detection errors by tuning the estimated trade-off of false positives and false negatives according to the perceived risk of the environment. This occurs via a change of the school's density and thereby its distance to criticality and \textit{not} primarily by being at the critical point and thus maximally distinguishing different inputs (number of initial startlers). 
The absence of a single optimal amplification scheme (often assumed to be exactly at the critical point) may be due to the high variability in the environment and also variability of initial responses to the same cue due to changing group compositions and inter-individual differences.

Based on our findings we suggest that the study of criticality in living systems, and in particular in animal collectives, focuses not only on the possible computational benefits but also on the potential costs of amplification of irrelevant fluctuations. By presenting a concrete example of a system that makes use of different distances to the critical point according to context, we emphasize the benefits of actively regulating
distance to criticality according to the environmental conditions or the complexity of the computational task at hand, as recently also discussed in the context of neuroscience \cite{Cramer2020}.
We further note that while investigation of neuronal cascades in vitro provide evidence for criticality \cite{Gross2021ManyCriticalStates,shew2011information,beggs2012being}, analysis of spike-avalanches in vivo suggests subcritical dynamics \cite{priesemann2014spike} in line with our findings. This points to the possibly higher importance of robustness relative to sensitivity in collective information processing.

\section*{Materials and Methods}
\subsection*{Experimental data}
This study uses experimental tracking data of schools of juvenile golden shiners, described in \cite{SosTwoBak19} and \cite{Rosenthal2015}, swimming freely in a white tank with shallow water, keeping them approximately two dimensional. Startling cascades were initiated by randomly occurring single startles (without provision of a stimulus or presence of a predator) and extracted from the tracking data. Previous work has found that these false alarms spread indistinguishably from startles that are initiated by a real fearful tactile stimulus \cite{Rosenthal2015}. From \cite{SosTwoBak19} we use data of groups before and after a natural alarm substance (Schreckstoff, a family of chondroitins released from injured fish skin, i.e. close to a successful predation event \cite{Mathuru2012}) was automatically sprayed on the water surface. For more details we refer to reader to the original studies \cite{Rosenthal2015,SosTwoBak19}.

\subsection*{Behavioral contagion model}
To simulate startle cascades we use a SIR type model of behavioral contagion based on \cite{dodds2004universal,dodds2005generalized} and previously used in \cite{SosTwoBak19}. An individual, represented by a node in an interaction network with edges $w_{ij}$, can be in one of three states: susceptible (swimming at base speed), active (startling) and recovered (having startled in the cascade already). At the beginning of the simulation we set $n\geq1$ individuals to the active state and we stop the simulation once no more active individuals remain. 

The individuals' internal dynamics are as follows: a susceptible individual $i$ receives stochastic activation cues of fixed size $d_\text{a}$ from active individuals $j$ with a rate $r_{ij}=\rho_\text{max}w_{ij}$ proportional to their link strength in the interaction network. Because the interaction networks are based on vision this means the individual is only influenced by those visible to it, i.e. those with $w_{ij}>0$. The individual internally integrates the stochastic time series of received cues, $d_{ij}(t)$, equally over its recent history of length $\tau_{m}$ (its memory time) to obtain its current cumulative activation
\begin{equation}
D_{i}(t)= \frac{1}{K_i}\sum_j \int_{t-\tau_m}^{t} d_{ij}(t') \, dt'~.
\end{equation}
Here $K_i$ is the number of $i$'s visible neighbors making this a fractional contagion process as supported by previous work \cite{Rosenthal2015}. Each individual has an internal response threshold $\theta$, that indicates the level of socially signaled risk it tolerates before startling itself. If the cumulative activation exceeds this threshold the individual becomes active. After a fixed activation time $\tau_\text{act}$ it recovers and stays that way until the end of the simulation. Response thresholds are drawn from a uniform distribution with minimum $\theta_{min}=0$ and maximum $\theta_\text{max}=2\bar{\theta}$ resulting in an average response threshold of $\bar{\theta}$ to account for stochasticity due to inaccessible internal states of individuals at the time of initial startle. Most parameters are fixed as in \cite{SosTwoBak19} ($\tau_\text{act}=0.5\textnormal{s},~\rho_\textnormal{max}=10^3 \textnormal{s}^{-1},~d_\text{a}=10^{-3}$) leaving the average response threshold $\bar{\theta}$ as a single free parameter describing the individual sensitivity to social cues.
Similar to \cite{SosTwoBak19} we fit the average response threshold to the experimental data using a maximum likelihood approach with 10,000 simulations of a cascade per network, initiated by the same individual as experimentally observed for this network (SI, Fig. \ref{fig:ellipse_validation} and table \ref{tab:LL_ellipse_fovea}).  We set the finite memory to $\tau_m$=1s as link weights $w_{ij}$ were obtained from the experimental data using a time window of $1$~second to detect startles following an initial startle event \cite{SosTwoBak19}, however, our general results are robust with respect to the choice of $\tau_m$ (see section \ref{sec:agentmem} of the SI).

\subsection*{Construction of interaction networks}
Interaction networks are based on experimentally observed response rates to a single initial random startle. Previous work found that the probability of individual $i$ to be the first responder to a startle of individual $j$ can be described as 
\begin{equation}
 w_{ij}=(1+\exp(-\beta_1-\beta_2\text{LMD}-\beta_3\text{RAA}))^{-1}
 \label{eq:links}
\end{equation}
with $\text{LMD}$ the log of the metric distance (in cm) between individual $i$ and $j$ and $\text{RAA}$ the ranked angular area of $j$ in the visual field of $i$ \cite{Rosenthal2015,SosTwoBak19}. Coefficients $\beta_i$ are obtained from a logistic regression of experimental observations of first responders and can be found in the SI (table \ref{tab:betas}).
For the construction of networks of different densities we rescaled the positions of the fish and recalculated metric distances and visual fields to obtain network links. Individuals were approximated as ellipses and a particle-based simulation of ellipse-shaped particles (based on \cite{Palachanis2015}) was used to shift and slightly turn individuals where they overlapped in order to ensure a 2 dimensional school. Details can be found in the SI, section \ref{sec:rescalednets}.

We validate the ellipse approximation by comparing the networks we obtain for the original (non-rescaled) density with the networks constructed in previous research \cite{Rosenthal2015} (using ray casting and body shape reconstruction from the tracking software). We compared the two sets based on network properties and the ability of the behavioral contagion model (see previous paragraph) to explain the observed data using these networks (SI, \ref{sec:ellipse_validation}) and found good agreement.

\subsection*{Sensitivity}
We simulate 1,000 startle cascades per experimental trial and rescaling factor by setting $n$ randomly chosen individuals to the active (startling) state and recording the number of individuals that startle before the cascade dies out. From these simulations we obtain an average cascade size for each trial, initial condition ($n$) and bins of median nearest neighbor distance by averaging over the simulation runs. Division by group size, $N$, yields the relative average cascade size. Subtracting the (relative) average cascade sizes following $n$ initial startles from that following $n+1$ initial startles gives the sensitivity shown in Fig. \ref{fig:sensitivity_and_netneighbors} of the SI, with the special case of $n=1$ shown in Fig. \ref{fig:sensitivity_1d}B.
Subtracting the relative average size of cascades triggered by one individual from that triggered by $n$ individuals yields the sensitivity shown in Fig. \ref{fig:sensitivity_and_netneighbors} of the SI. The qualitative result does not depend on choosing initiators at random, but holds for choosing them to be network neighbors as well, see SI Fig. \ref{fig:sensitivity_and_netneighbors}.

\subsection*{Branching ratio calculation}
A useful measure to describe the aggregate effects of a local perturbation in
a contagion model is the average branching ratio, which answers the question:
In a completely susceptible (quiescent) system in which a single individual $j$ 
becomes active, what is the average number of other individuals $i$ that become
active due to the direct influence of individual $j$?  This branching ratio $b$
describes how small initial startling cascades, on average, tend to grow or shrink.  
For large $N$, it also defines a transition between two distinct aggregate behaviors:
when $b < 1$, startling cascades tend to die out and only affect a small fraction of the whole group, and when $b > 1$, cascades tend to spread through the entire group. To calculate $b$, we note that the average total additional activation received by
individual $i$ due to $j$ startling is \cite{SosTwoBak19}
\begin{equation}
    \Delta D_i = d_\text{a} \rho_\mathrm{max} \tau_\mathrm{act} \frac{w_{ij}}{K_i},
\end{equation}
where $d_\text{a}$ is the activation cue intensity, $\rho_\mathrm{max}$ is the maximum rate of
receiving cues, $\tau_\mathrm{act}$ is the activation duration, $K_i$ is the number
of $i$'s visible neighbors,
and $w_{ij}$ is the probability from the logistic regression model 
that $i$ is a first responder given that individual $j$ initially startled.
As in Ref.~\cite{SosTwoBak19}, we will discretize time in 1\,ms increments and
set the arbitrary scale of cue intensities $d_\text{a}$ such that $d_\text{a} \rho_\mathrm{max} = 1$, producing
\begin{equation}
    \Delta D_i = \tau_\mathrm{act} \frac{w_{ij}}{K_i}.
\end{equation}
We translate this additional activation to individual $i$ into the average number of additional startles produced by calculating the probability that this causes $i$'s state to go above threshold.  This is equal to $\Delta D_i$ times the probability density of $i$'s threshold being between $D_i$ and $D_i + \Delta D_i$.  
Because we use a 
constant probability density for thresholds 
$p(\theta_i) = \theta_\mathrm{max}^{-1}$, and  
given $\Delta D_i < \theta_\mathrm{max}$, the probability of
$i$ startling due to $j$ startling is
\begin{equation}
    p_{ij} = \Delta D_i \theta_\mathrm{max}^{-1} 
           = \frac{\tau_\mathrm{act}}{\theta_\mathrm{max}} 
             \frac{w_{ij}}{K_i}.
\end{equation}
Then the average number of additional individuals startled due
to $j$'s startle is our branching ratio 
\begin{equation}
b_j = \sum_i p_{ij}=\frac{\tau_\mathrm{act}}{\theta_\mathrm{max}}\sum_i \frac{w_{ij}}{K_i}
\end{equation}
and the median branching ratio, given that the cascade starts
with a random individual $j$, is the measure used in this study as an analytical measure of criticality.

\subsection*{Relative individual payoff}
Assuming individual pay-offs due to different costs of false and true positives and negatives, we have derived the relative payoff, $\psi$, as a function of median nearest neighbor distance (NND) to
\begin{equation}
    \psi(\text{NND})=-\frac{1}{\psi_0}\left[p_\text{fp}(\text{NND})\cdot r+p_\text{fn}(\text{NND})\right]~,
    \label{eq:payoff}
\end{equation}
(see detailed derivation in section \ref{sec:SI_payoff_derivation} of the SI). In Eq. \ref{eq:payoff}, the relative noise cost, $r=\frac{\rho_n}{\rho_p}\left(\frac{c_\text{fp}-c_\text{tn}}{c_\text{fn}-c_\text{tp}}\right)$, characterizes the environment via the costs associated to each event type (false and true positive, $c_\text{fp}$ and $c_\text{tp}$, as well as false and true negative, $c_\text{fn}$ and $c_\text{tn}$), and the rate at which predator cues, $\rho_ p$, and noise cues, $\rho_n$, appear. 
The conditional probabilities of the two types of error, startling ($S$) in response to noise (false positives, $p_\text{fp}$), or not startling ($\bar{S}$) in response to a predator cue (false negatives, $p_\text{fn}$), are determined through simulations of average cascade sizes. Depending on the type of error, cascades are initiated by different numbers, $N_\text{init}$, of randomly selected individuals as follows
\begin{align}
    \begin{aligned}
    p_\text{fp}(\text{NND})&=p(S|N_\text{init}=1,\text{NND}),\\
    p_ {fn}(\text{NND})&=p(\bar{S}|N_\text{init}=p_\text{detect}N_\text{vis}(\text{NND}),\text{NND}).
    \end{aligned}
\end{align}
Referring to experimental observations we use one initial startler to simulate false positives. False negatives are simulated as the average fraction of a school that is not part of a cascade initiated by $p_\text{detect}N_\text{vis}(\text{NND})$ individuals. Here, $N_\text{vis}(\text{NND})$ is the average number of individuals that can see a predator at distance $d_{pred}=$10\,BL from the school boundary with a maximal detection range of $d_\text{max}$=40\,BL and $p_\text{detect}=0.1$ characterizes the individual responsiveness to a predator cue. Average cascade sizes for non-integer numbers of initial startles are obtained by proportionally combining results for the two integers closest to the desired value. The relative payoff is rescaled by the costs for an individual in an infinitely dilute school,  
\begin{equation}
    \psi_0=\lim_{\text{NND} \to \infty}[p_\text{fp}(\text{NND})\cdot r+p_\text{fn}(\text{NND})]=1-\frac{p_\text{detect}-r}{N},
\end{equation}
to highlight the influence of schooling. Qualitatively our results are independent of the exact choice of $d_{pred}$, $d_\text{max}$ and $p_\text{detect}$ (see SI, figure  \ref{fig:SI_visualfraction} and \ref{fig:vis_detection}).

\section*{Acknowledgements}
\footnotesize{
W.P. and P.R. were funded by the Deutsche Forschungsgemeinschaft (DFG) (German Research Foundation), Grant RO47766/2-1 . P.R. acknowledges funding by the DFG under Germany’s Excellence Strategy–EXC 2002/1 “Science of Intelligence”–Project 390523135. 
B.C.D. was supported by a fellowship at the Wissenschaftskolleg zu Berlin and by the ASU–SFI Center for Biosocial Complex Systems. M.M.G.S. was supported by an NSF Graduate Research Fellowship. C.R.T. was supported by a MindCORE (Center for Outreach, Research, and Education) Postdoctoral Fellowship. 
I.D.C. acknowledges support from the NSF (IOS-1355061), the Office of Naval Research grant (ONR, N00014-19-1-2556), the Struktur- und Innovationsfunds für die Forschung of the State of Baden-Württemberg, the Deutsche Forschungsgemeinschaft (DFG, German Research Foundation) under Germany’s Excellence Strategy–EXC 2117-422037984, and the Max Planck Society.}
\bibliographystyle{unsrt}  
\footnotesize{
\bibliography{ms}
}
\newpage














\renewcommand\thesection{\arabic{section}}

\newcommand{\mean}[1]{\left < #1 \right >}
\newcommand{\abs}[1]{\left | #1 \right |}
\renewcommand{\vec}[1]{\mathbf{#1}}
\newcommand{\fourier}[2]{{\hat{#1}}_{#2}}
\newcommand{\nabcp}{\underset{\sim}{\nabla}}
\newcommand{\red}[1]{\textcolor{BrickRed}{#1}}
\newcommand{\green}[1]{\textcolor{ForestGreen}{#1}}
\newcommand{\blue}[1]{\textcolor{NavyBlue}{#1}}

\renewcommand{\thepage}{\arabic{page}} 
\renewcommand{\theequation}{S\arabic{equation}} 
\renewcommand{\thesection}{\arabic{section}}  
\renewcommand{\thetable}{S\arabic{table}}  
\renewcommand{\thefigure}{S\arabic{figure}}

\newcommand{\eqrefit}[1]{\textit{(\ref{#1})}}

\newcommand{\refSIsec}[1]{SI Section \ref{#1}}


\begin{flushleft}
{\Large
\textbf{SUPPLEMENTARY INFORMATION}
\ \vspace{8pt} \\
\textbf{Subcritical escape waves in schooling fish}
}
\ \vspace{8pt} \\

\noindent\textbf{Winnie Poel$^{a,b}$,
Bryan C. Daniels$^{c}$,
Matthew M. G. Sosna$^{d}$,
Colin R. Twomey$^{e}$,
Simon P. Leblanc$^{d,f}$,
Iain D. Couzin$^{g,h,j}$
Pawel Romanczuk$^{a,b,*}$
}
\vspace{8pt} \\
$^{a}$Institute for Theoretical Biology, Department of Biology, Humboldt Universit{\"a}t zu Berlin, D-10099 Berlin, Germany \\
$^{b}$Bernstein Center for Computational Neuroscience, Berlin, D-10115 Berlin, Germany \\
$^{c}$School for Complex Adaptive Systems, Arizona State University, Tempe, AZ 85287 \\ 
$^{d}$Department of Ecology and Evolutionary Biology, Princeton University, Princeton, NJ 08544 \\
$^{e}$Department of Biology, University of Pennsylvania, Philadelphia, PA, 19104 \\ 
$^{f}$Blend Labs, San Francisco, CA 94108\\
$^{g}$Department of Collective Behaviour, Max Planck Institute of Animal Behavior, D-78547 Konstanz, Germany \\
$^{h}$Department of Biology, University of Konstanz, D-78547 Konstanz, Germany \\
$^{j}$Centre for the Advanced Study of Collective Behaviour, University of Konstanz, D-78547 Konstanz, Germany \\
$\ast$ E-mail: pawel.romanczuk@hu-berlin.de
\end{flushleft}

\renewcommand{\thepage}{\arabic{page}} 
\renewcommand{\theequation}{S\arabic{equation}} 
\renewcommand{\thesection}{\arabic{section}}  
\renewcommand{\thetable}{S\arabic{table}}  
\renewcommand{\thefigure}{S\arabic{figure}}



\section{Construction of rescaled interaction networks}
\label{sec:rescalednets}
This section describes how the interaction networks for densities that were not observed in experiments were constructed. Briefly said, we 
\begin{itemize}
    \item first rescaled the position data,
    \item then approximated each individual fish by an ellipse,
    \item used an active particle simulation to ensure that when placed at the rescaled positions using the original orientations,
    the ellipses would relax into a non-overlapping configuration,
    \item then calculated the visual field of each ellipse using an analytically derived formula and afterwards accounting for occlusions using an algorithm based on the analytic results to determine the angle which one ellipse occupies in the visual field of another ellipse,
    \item before finally generating the interaction network based on the calculated values of log metric distance and ranked angular area.
\end{itemize}
We conclude this section with a comparison of the properties of the thus constructed networks with the networks previous studies \cite{SosTwoBak19, Rosenthal2015} constructed based on bodyshape reconstruction from tracking software and raycasting. A comparison of the two types of networks based on their ability to describe the observed cascade size distributions when used together with the behavioral contagion process can be found in the next section.
\subsection{Rescaling position data}
The position data was rescaled with the following range of factors: $d\in [0.2,0.3,\dots,2.9,3.0]$ setting \begin{equation}\begin{aligned}
x_{\text{rescaled}}=&d\cdot x_{\text{exp}}\\
y_{\text{rescaled}}=&d\cdot y_{\text{exp}}~.
\end{aligned}\end{equation}
\subsection{Ellipse approximation}
To approximate a single fish we used an ellipse with its center at $(x_0,y_0)$, semi-major axis of length $a$ and semi-minor axis of length $b$. The semi-major axis is rotated by the fish's orientation, $\phi$, from the x-axis. We call $w=b/a$ the aspect ratio of the ellipse. The ellipse has a single eye that is positioned on the semi-major axis at a distance $l$ from the center of the ellipse, where $l=1$ refers to the front, $l=0$ the center and $l=-1$ to the back of the ellipse (see Fig. \ref{fig:sketch_ellipse}). We chose $w=0.14$ and $l=0.9$ for all fish in agreement with the average value obtained from tracking software data where we used the ratio of detected inter eye distance and bodylength of a fish as $w$ and the position of the eye along the detected midline to determine $l$, see Fig. \ref{fig:hist_w}. Because the tracking software saved the edge of the head of the fish as its position $(x_{\text{exp}}, y_\text{exp})$ we positioned the ellipse at
\begin{equation}\begin{aligned}
x_0&=x_\text{exp}-\frac{a}{2}\cos{\phi}~,\\
y_0&=y_\text{exp}-\frac{a}{2}\sin{\phi}~.
\end{aligned}
\end{equation}

\subsection{Active particle simulation to eliminate overlaps}
Since we assume our school of fish to be 2 dimensional there is an upper limit to the density we can achieve by rescaling (see also section \ref{sec:densitylimit} of the SI). We acknowledge this by ensuring that no two fish overlap in the rescaled position data using an active particle simulation based on the code provided within \cite{Palachanis2015}. Ellipses are placed at the rescaled positions at $t=0$ with velocities $v_i(t=0)=0$. Their velocities $\vec{v}_i(t)$ change based on their positions $\vec{x}_i(t)=(x_i(t),y_i(t))^T$ according to 
\begin{equation}
    \begin{aligned}
         \frac{d\vec{v}_i(t)}{dt}&=\frac{1}{m_i}\left(-\alpha\vec{v}_i(t)+\sum_{j\neq i}^j\frac{\vec{x}_i-\vec{x}_j}{|\vec{x}_i(t)-\vec{x}_j(t)|}F_{ij}(t)\right)\\
        F_{ij}&= \lambda A_{ij}(t)
    \end{aligned}
\end{equation}
where $A_{ij}(t)$ is the overlap area of the ellipse $i$ and ellipse $j$ at time $t$, $m_i=m=1$ is the mass of the ellipse, $\alpha$ the damping parameter and $\lambda$ a constant model parameter. We use $\alpha=0.2$ and $\lambda=0.05$ and stop the simulation as soon as no more overlaps are detected. To speed up the relaxation into a non-overlapping state, we used a second larger ellipse (factor 1.1) to determine the repulsion area $A_{ij}$ and stopped when the original size ellipses were no longer overlapping, see Fig. \ref{fig:overlap}.

\subsection{Analytical calculation of visual field of ellipse}
\begin{figure}
\centering
\includegraphics[scale=.7]{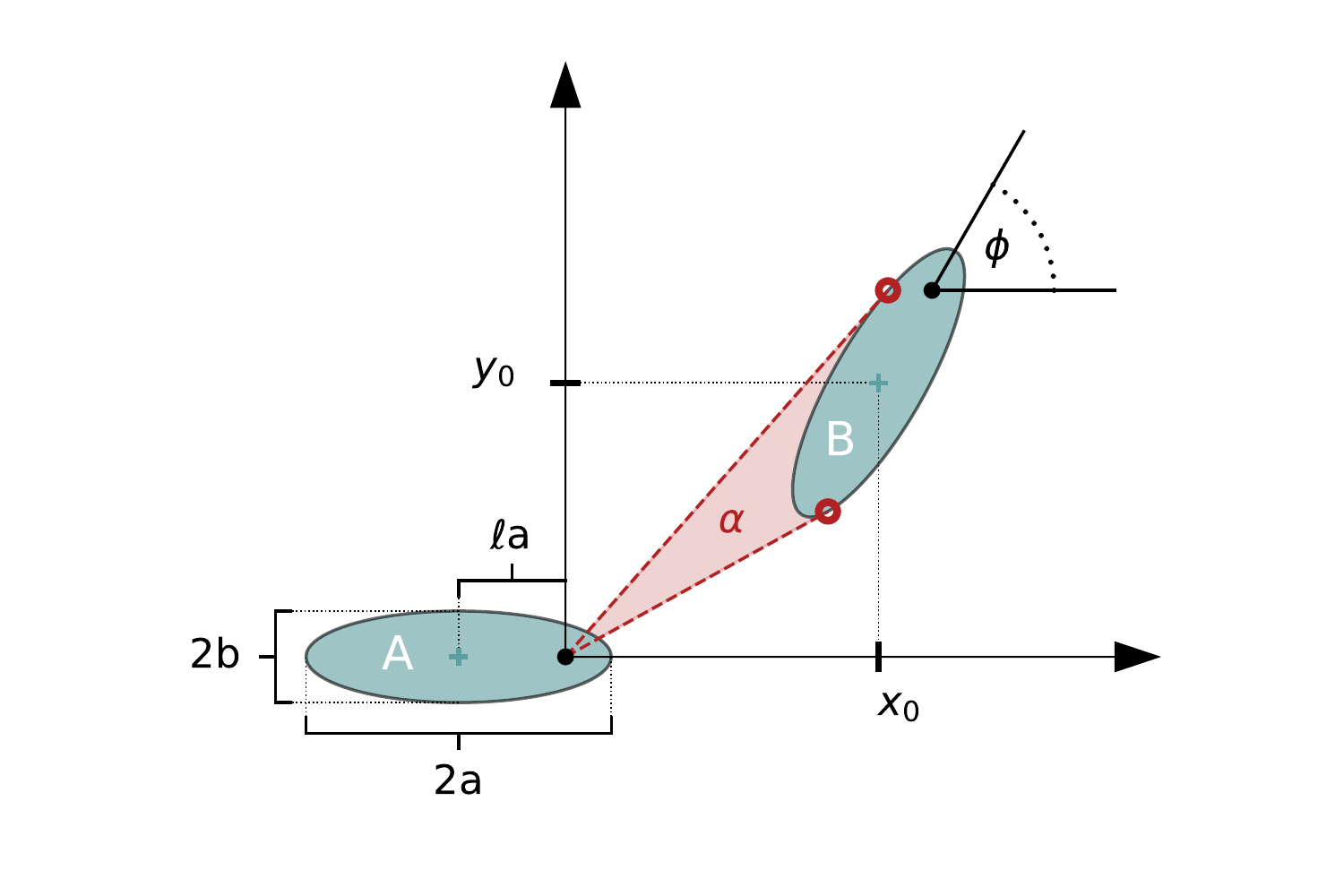}
\caption{Illustration of the ellipses used to approximate an individual fish and variables used in the analytical calculation of the ellipse's visual field. The visual angle, $\alpha$, of ellipse B in the visual field of ellipse A, is colored red and given by the angle between the two tangent lines (dashed, red) which intersect ellipse B in a single point (red) each.}
\label{fig:sketch_ellipse}
\end{figure}

\begin{figure}
\centering
\includegraphics[scale=0.7]{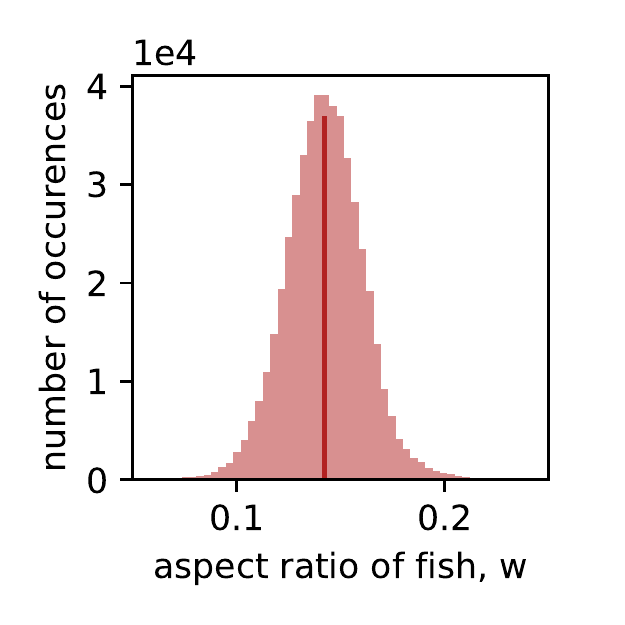}
\includegraphics[scale=0.7]{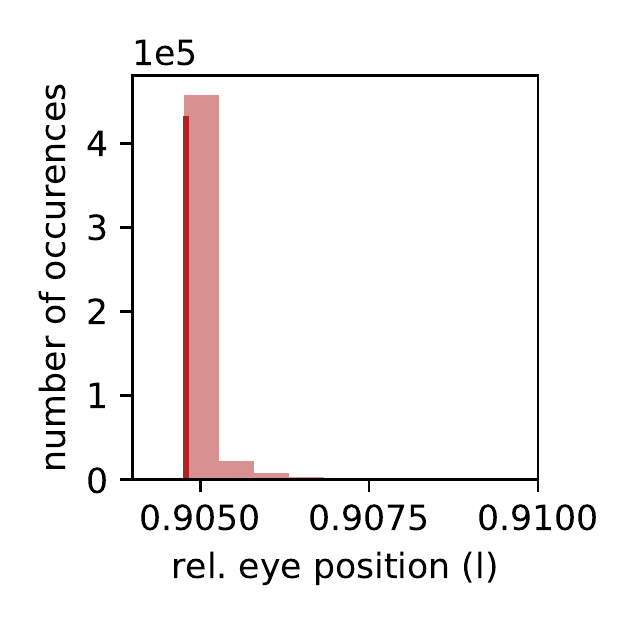}
\caption{Determining aspect ratio $w$ and eye position $l$ used to approximate fish by ellipses from tracking data. This histogram is based on a video from the experiments performed in \cite{Rosenthal2015}.}
\label{fig:hist_w}
\end{figure}

\begin{figure}[htp]
    \centering
    \includegraphics[scale=0.35]{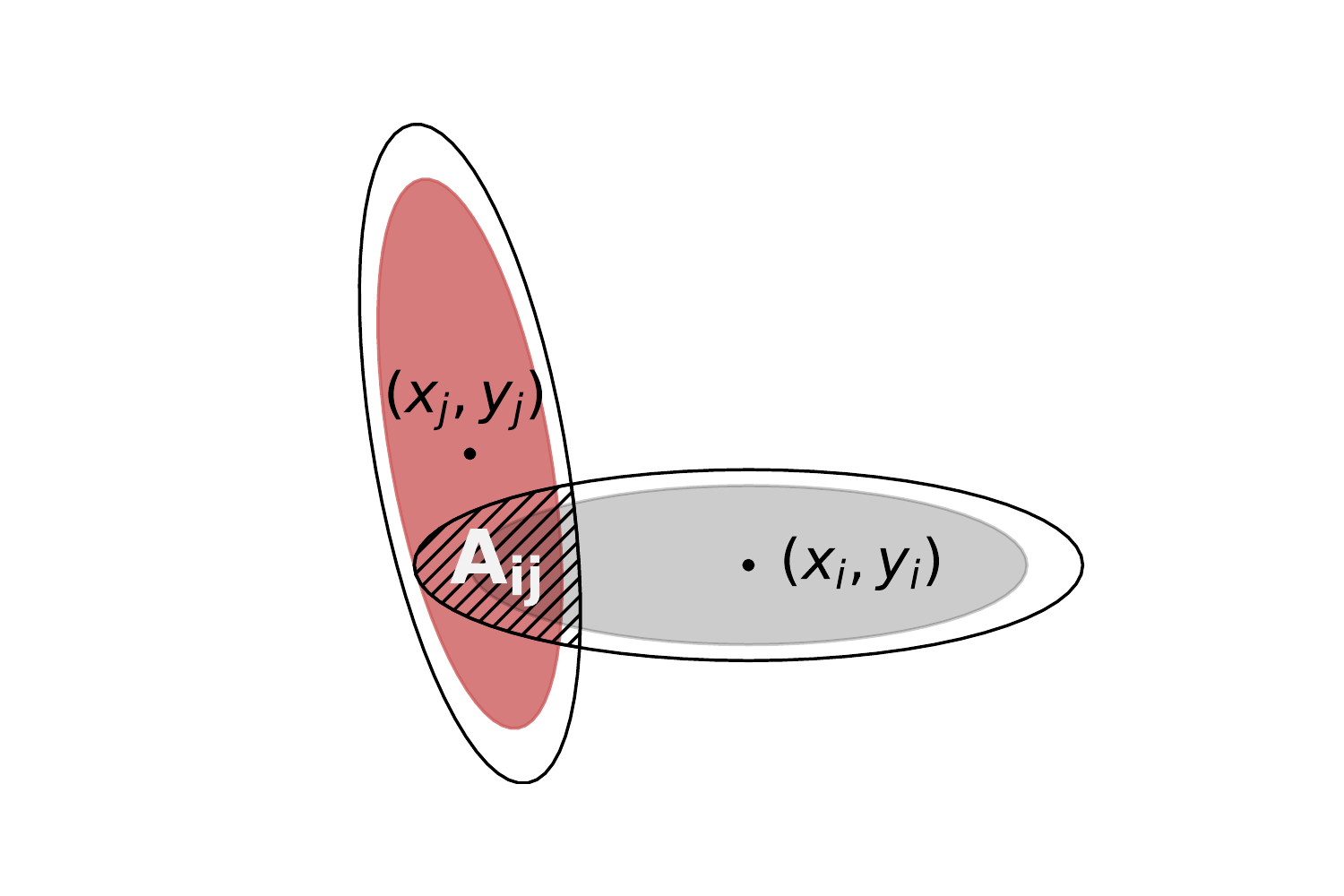}
    \caption{Illustration of the ellipse interactions in the active particle simulation. Ellipses repel each other based on their overlap area (hatched). To speed up the simulation and avoid very small forces towards the end we use a larger ellipse (rescaled by a factor 1.1, black lines) to calculate overlap area and stopped the simulation when the original sized ellipses (colored areas) do not overlap anymore.}
    \label{fig:overlap}
\end{figure}

In order to determine the interaction network of the ellipses we need a matrix of ranked angular areas and distances for all pairs of ellipses (see eq. \ref{eq:pij}). The ranked angular areas were determined based on the analytical calculation of the angular area that an ellipse at a relative position $(x_0,y_0)$, semi-major axis of length $a$ rotated by $\phi$ from the x-axis and semi-minor axis of length $b$ has for an observer sitting at the origin. We assume the observer has 360$^\circ$ vision.
The observed ellipse is given by:
\begin{equation}
\begin{aligned}
 \left(\begin{array}{c}x\\y\end{array}\right)&=\left(\begin{array}{c} x_0+a\cos\psi\cos\phi+b\sin\psi\sin\phi\\
						     y_0+a\cos\psi\sin\phi-b\sin\psi\cos\phi\end{array}\right)\\
						     \text{with}&\quad0\leq \psi<2\pi~.
\label{eq:ellipse}	
\end{aligned}
\end{equation}
To describe the fish we use the aspect ratio $w$ and polar coordinates $r, \theta$
\begin{equation}
  \begin{aligned}
  a&=1/2\\
  b&=w/2\\
  x_0&=r\cos\theta\\
  y_0&=r\sin\theta~.
  \end{aligned}
\end{equation}
We can then calculate the gradient of the ellipse as
\begin{equation}
\begin{aligned}
 \frac{dy}{dx}&=\frac{\sin\phi\sin\psi+w\cos\phi\cos\psi}{\cos\phi\sin\psi-w\sin\phi\cos\psi} ~.
 \end{aligned}
\end{equation}
A tangent line to the ellipse is given by
\begin{equation}
 y=\left.\frac{dy}{dx}\right|_s(x-x_s)+y_s
\end{equation}
where $(x_s,y_s)$ is the tangent point on the ellipse (see Fig. \ref{fig:sketch_ellipse}).

Since we place our observer at the origin, the tangents we need in order to determine the visual field have to pass through this point and thus it needs to hold that
\begin{equation}
\begin{aligned}
0&=\left.\frac{dy}{dx}\right|_s(0-x_s)+y_s\\
0&=-\left.\frac{dy/d\psi}{dx/d\psi}\right|_sx_s+y_s\\
0&=-(dy/d\psi)|_sx_s+(dx/d\psi)|_sy_s
\end{aligned}
\end{equation}
which can be written as
\begin{equation}
\begin{aligned}
 0&=\left|\begin{array}{cc} x_s & dx/d\psi|_s\\
       y_s & dy/d\psi|_s
      \end{array}\right|\\[0.5cm]
       x_s&=r\cos\theta+\frac{\cos\phi\cos\psi+w\sin\phi\sin\psi}{2}\\
       \frac{dx}{d\psi}|_s&=\frac{-\cos\phi\sin\psi+w\sin\phi\cos\psi}{2}\\
       y_s&=r\sin\theta+\frac{\sin\phi\cos\psi-w\cos\phi\sin\psi}{2}\\
       \frac{dy}{d\psi}|_s&=-\frac{\sin\phi\sin\psi-w\cos\phi\cos\psi}{2}
\end{aligned}
\end{equation}
where $||$ is the determinant.
Solving for $\psi$ yields

\begin{equation}
\begin{aligned}
\psi_{\pm}=&\pm2\tan^{-1}\left(\frac{\gamma\mp r\sin(\theta - \phi)}{\beta}\right)\\
\beta=& w(2r\cos(\theta - \phi) - 1)\\
 \gamma=&\left(-w^2 + 2r^2\left((1 + w^2)\right.\right. \\
 &\left.\left.+ (w^2 - 1)\cos(2(\theta-\phi))\right)\right)^{1/2}~.
\end{aligned}
\label{eq:psi_tps}
\end{equation}
Inserting \eqref{eq:psi_tps} into \eqref{eq:ellipse} returns the tangent points $(x_{x1}, y_{s1})$ and $(x_{x2}, y_{s2})$ whose polar angles, $\theta_{s1}$ and $\theta_{s1}$, determine the angular area of the ellipse, $\alpha=\text{min}(|\theta_{1s}-\theta_{2s}+n\pi|, n\in\mathbb{N})$ (here we use the fact that the angular area needs to be smaller than $\pi$ because the ellipses aren't allowed to overlap).

Occlusions of individual $j$ in the visual field of $i$ by all other individuals $k\neq j$ are then determined by an algorithm using intersections of rays originating from the eye of the focal individual $i$ and going through the tangent points (as perceived by $i$) on $j$ with the outlines of ellipses $k$. To determine if $j$ or $k$ is visible to $i$ in the occluded area, intersection of ellipse outlines with the angle bisectors of the rays described above are also considered. 


\subsection{Network construction}
    As described in the methods section of the main paper, the weight of a link (from $i$ to $j$) is given by:
    \begin{equation}
     w_{ij}=(1+\exp(-\beta_1-\beta_2LMD-\beta_3RAA))^{-1}~.
     \label{eq:pij}
    \end{equation}

    $LMD$ is the log of the metric distance between ellipse $i$ and $j$ and $RAA$ is the ranked angular area of $j$ in the visual field of ellipse $i$. Coefficients $\beta_i$ were obtained from a logistic regression of experimental observations of first responders and can be found in table \ref{tab:betas}. Data from before and after exposure to Schreckstoff were fitted with one set of coefficients in accordance with \cite{SosTwoBak19} who did not find any significant change from `Baseline' to `Alarmed' state.

\subsection{Density dependence of network measures}
\begin{figure}
    \centering
    \includegraphics[scale=.7]{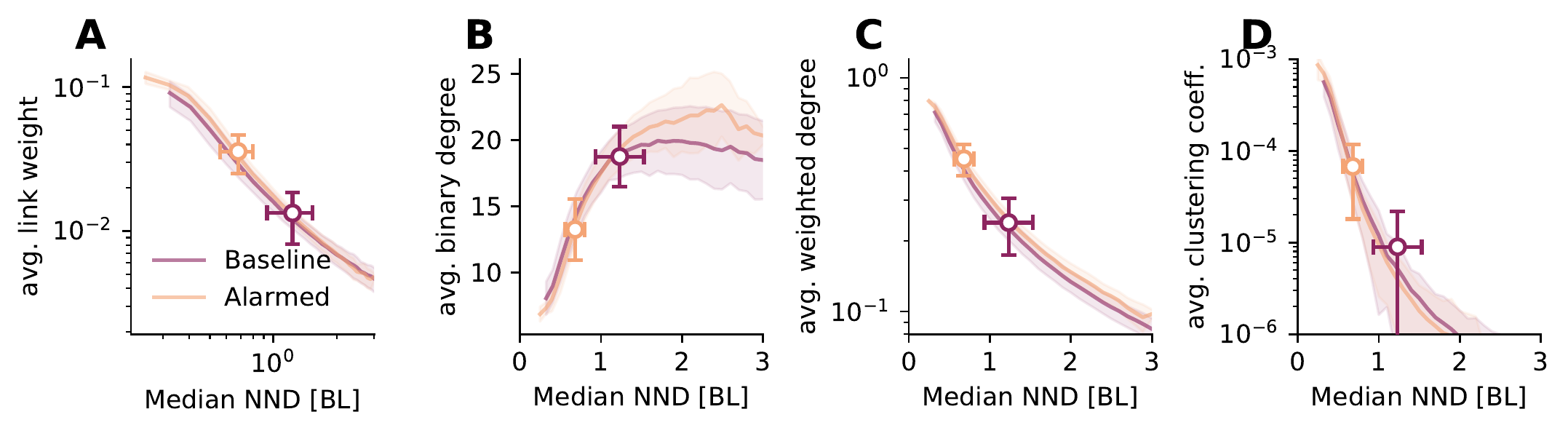}
    \caption{Network properties as a function of median nearest neighbor distance (NND) for the different datasets (lines represent averages over networks obtained from different experimental trials, shaded areas the standard deviation between networks). Dots with errorbars represent averages $\pm$ one standard deviation of the mean over the original scale experimental trials. Average link weight (A) and average weighted degree (C) decrease  with NND. Average number of network neighbors (B) peaks at intermediate densities. At low NND occlusions limit the number of neighbors, at high NND the visual threshold leads to a decrease in neighbors. The clustering coefficient (D), which was shown to predict cascade sizes \cite{Rosenthal2015}, decreases with NND.}
    \label{fig:netprops}
\end{figure}

Fig. \ref{fig:netprops} shows different network properties plotted against median nearest neighbor distance (NND) in body lengths. For more details refer to the figure caption.

\section{Model Calibration}
        As described in the methods section of the main paper, the behavioral contagion model has one free parameter, the response threshold, that corresponds to an individual sensitivity to social cues. We calibrate the model separately for each dataset using the experimentally observed cascade size distributions following a maximum likelihood approach as in \cite{SosTwoBak19}. This allows us to determine which value of the parameter response threshold best describes the observed data.
    
        Each startle cascade observed in the experiments corresponds to one interaction network based on the positions and visual fields of all individuals at the time just before the initial startle. In our model we artificially initiate a startling cascade on this network by setting the node corresponding to the experimentally observed initial startler to the active state at $t=0$ and recording the resulting cascade size (i.e. the number of individuals, including the initial startler, changing their state to active before the cascade dies out and no active individuals remain). This is done 10000 times for each interaction network and for a range of response threshold values. For each set of networks (corresponding to one of the experimental conditions) we combine the cascades simulated for all interaction networks of this set and build a distribution of cascade sizes for each response threshold. These probability distributions combined with the experimentally observed cascade size distribution allow us to determine which response threshold is most likely given our observed data (i.e. has the maximum likelihood). The result can be seen in Fig. \ref{fig:ellipse_validation} for both the networks constructed using ellipses to approximate body shape and visual field (solid lines) as well as networks based on the tracking software described in \cite{SosTwoBak19,Rosenthal2015}. .
    
        Fig. \ref{fig:modelfitoneset} shows the calibrated model and the experimental data it describes. 

\section{Model assumptions and limitations}
\subsection{Assessing the quality of ellipse based networks}
    \label{sec:ellipse_validation}
    To see the effects of our approximation of bodyshape and field of view using ellipses and analytical calculation we first compare the resulting networks via their link weight distribution, as well as their (weighted and binary) degree distribution. Then, to see the influence our approximation may have on the dynamics of the complex contagion process and the model's ability to describe the observed data, we fit the single free parameter of the model, the response threshold, once for the (original scale) ellipse networks and once for the networks constructed using the ranked angular area determined through ray casting. 
    
     The distributions of network properties for both types of networks are shown in Fig. \ref{fig:comparing_ellipse_network_props} where the ellipse networks are drawn as a line and the ray casting networks as a shaded area.  The histograms show (from left to right): weighted degree, binary degree, link weights, link distance (metric distance between two fish connected by a link). Each row corresponds to one of the data sets, top: 40 fish `Baseline', middle: 40 fish `Alarmed', bottom: 150 fish. Overall, the distributions are generally in good agreement. The remaining differences between them are due to the fact, that for the ellipse networks all overlaps of individuals were eliminated to ensure a completely 2 dimensional school whereas in the experiments the shallow water still allowed the fish to occasionally cross path and appear stacked on top of each other in the data. The elimination of these overlaps by shifting the individuals apart leads to on average larger inter-individual distances in the ellipse networks, as seen in the right column of Fig. \ref{fig:comparing_ellipse_network_props}. Additionally, putting these originally stacked individuals next to each other in 2 dimension now means that they occlude each other's field of view, on average leading to fewer visible neighbors and thus fewer links per individual. This results in a shift in the weighted and binary degree distributions (two left columns, fig. \ref{fig:comparing_ellipse_network_props}) while the distribution of link weights does not change (second right column). The effect is especially prominent for the high density case (40 fish `Alarmed', second row) in which overlaps where most frequent.
     
     The results of the maximum likelihood fitting of the response threshold for both network constructions can be found in Fig. \ref{fig:ellipse_validation}. A plot of the experimental data with the cascade size distribution produced by the fitted models is shown in Fig. \ref{fig:modelfitoneset}.

\subsection{Density limit}
\label{sec:densitylimit}
    The regulation of group density has a naturally occurring limit towards high density because of the physical bodies of the fish. We need to consider this as a limit to the group's ability to modulate responsiveness solely via group density. When taken to the extreme this limit becomes the problem of closely packing ellipses of aspect ratio 0.14, but this limit certainly is not of biological relevance, especially in our model which is based on visual interactions. Instead we take a look at all the configurations we find in our original data (no rescaling) after ensuring that all overlaps have been resolved via the active particle simulation. The lower limit of density (measured by median nearest neighbor distance of the group) occurring in our experiments gives us an indication to what densities are possible as a group median, namely (40 fish: 0.44, 150 fish: 0.48), while individual fish can get closer (40 fish: 0.16, 150 fish: 0.15).
    Additionally, we constructed the density dependence of average number of physical contacts per individual from all the rescaled and original data we have (Fig. \ref{fig:touchpoints}). When a large fraction of the  ellipses start touching the limit of our vision based model is most likely reached as interaction can no longer be viewed as purely visual but become governed by physical forces and other sensory inputs. 
\subsection{Two dimensional schools}
\label{sec:SI_3d}
As in \cite{Rosenthal2015} and \cite{SosTwoBak19} we assume schools to be approximately 2 dimensional. As pointed out in the previous section, this yields a lower limit for the nearest neighbor distance (NND) of a school and by this an upper limit for the branching ratio that can be reached without an additional change in individual response threshold. In the main text and Fig. \ref{fig:branching} we find that this density limitation does not allow the schools to cross the critical manifold by a change of NND alone. Here, we make a simple argument to explain why we expect the maximum branching ratio that could be reached in a 3 dimensional school via a change of NND to be similar to that in 2 dimensions. 

As described in the methods section of the main paper, the branching ratio is calculated as  
\begin{equation}
    b_{j} = \frac{\tau_\mathrm{act}}{\theta_\mathrm{max}}\sum_i 
             \frac{w_{ij}}{K_i}.
             \label{eq:branching_SI}
\end{equation}
The branching ratio of an individual is thus maximized when the ratio of the weight of its incoming networks links, $w_j=\sum_iw_{ij}$ and the absolute number of its incoming network links $K_j$ is maximized. This is the case, when the average strength of a link in the network is largest, which is the case for small NND because of the functional dependence of $w_{ij}$ on interindividual distance. 

Thus because of the fractional contagion process and the resulting $K_i$ in the denominator of equation \eqref{eq:branching_SI}, it is the average link strength determining the maximal branching ratio at high densities and not the total weight of incoming networks links. For a densely packed configuration of ellipses, we expect the average link strength (corresponding to the links to the nearest neighbors) to be similar in 2 or 3 dimensions and thus the median branching ratio should be similar. A more detailed analysis would require the reconstruction of visual fields in 3 dimensions and is beyond the scope of this paper.

\subsection{Visual threshold}
\label{sec:visuallimits}
Because of limitations of the visual sensory perception and cognitive limitations it seems reasonable to assume that the angular area an individual $i$ occupies in the visual field of individual $j$, $\alpha_{ij}$ must be larger than a certain value, $\alpha>\alpha_{min}$ in order for individual $i$ to be seen by and influence the behavior of individual $j$. In the main text we use $\alpha_{min}=0.02$. Here, we explore further values. We construct networks and fit the individual response threshold of the model for each threshold value. Table \ref{tab:visthresh} gives these fitted individual threshold parameters. When larger thresholds are used, the average response threshold best describing the experimental data increases. The increased visual threshold lowers the average degree of a node, $K_i$, and thus because received cues are weighted with $1/K_i$ (see methods section of main text), received cue intensity becomes larger on average. To compensate for this effect, the average response threshold must be increased. Fig. \ref{fig:branching_visthresh} shows the dependence of the median branching ratio for the different visual thresholds. At high densities networks are similar for all thresholds because basically all neighbors are very close by and occupy a large angular area in the visual field and smaller (further away) individuals are blocked from view. The increased average response thresholds for larger visual thresholds then lead to a smaller branching ratio at high densities. Our main finding that observed schools are subcritical is not changed by this and neither is the general form of the relative individual payoff.

    \subsection{Choosing initiators}
    As seen in Fig. \ref{fig:sensitivity_and_netneighbors} our results do not change qualitatively when choosing casacade initiators as network neighbors instead of randomly as in the main text.

\subsection{Choice of agent memory}
\label{sec:agentmem}
For our study we have differed from \cite{SosTwoBak19} in our choice of agent memory by choosing $\tau_{mem}=1$ instead of $\tau_{mem}=2$. Fig. \ref{fig:agentmem} shows that our results do not critically depend on this choice. 

\subsection{Larger group size}
\label{sec:rosenthal}
\begin{figure}
        \centering
        \includegraphics[scale=0.6]{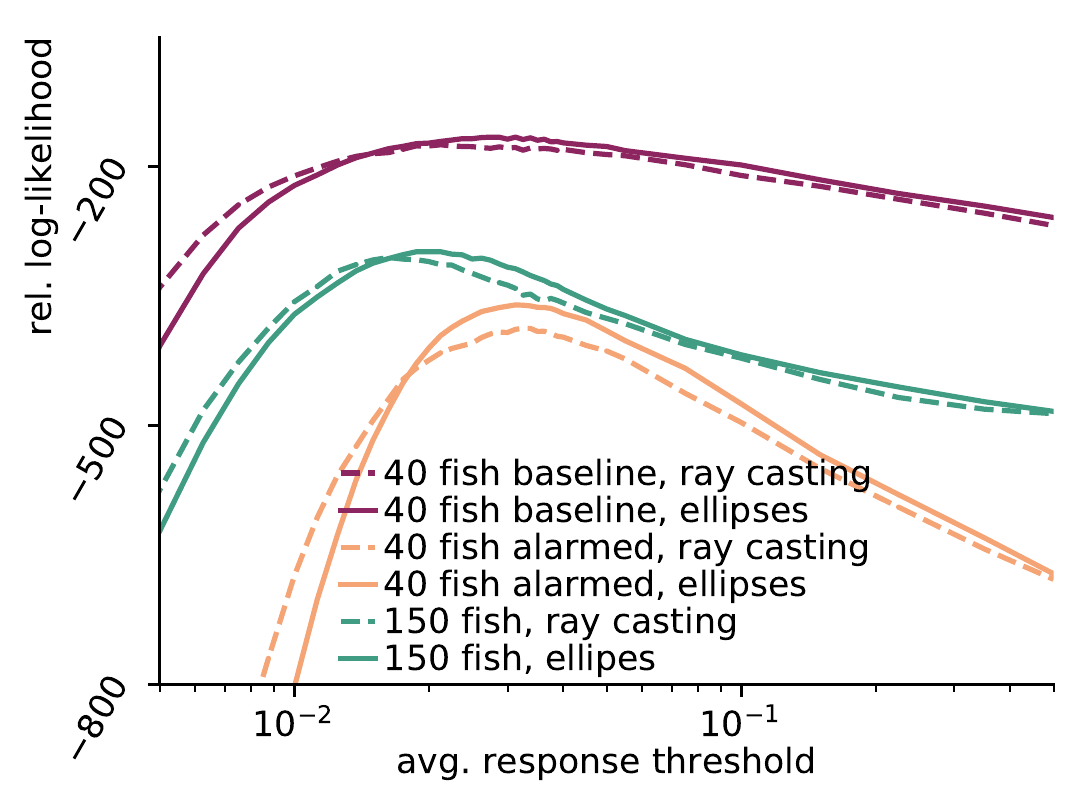}
        \includegraphics[scale=0.6]{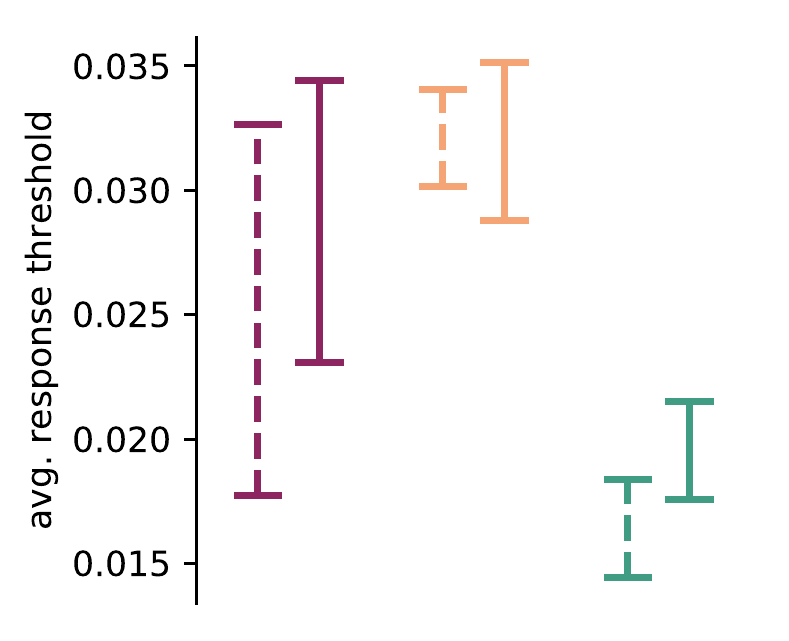}
        \caption{Calibration of the model: the average response threshold is chosen to maximize the rel. log-likelihood.  Left: rel. log-likelihood, right:  credible intervals for the best fit parameter. For details refer to table \ref{tab:LL_ellipse_fovea}. While in agreement with \cite{SosTwoBak19} the thresholds for the two experimental conditions of the groups of 40 fish have overlapping credible intervals, the group of 150 fish is best described by a lower threshold. This difference can be due to any of the differences in experimental setup and procedure and is in need of further research exploring this systematically. The ellipse based (dashed line) and the ray casting based (solid line) networks perform comparably well in describing the experimental data.}
        \label{fig:ellipse_validation}
    \end{figure}

\begin{figure}
        \centering
        \includegraphics[scale=.65]{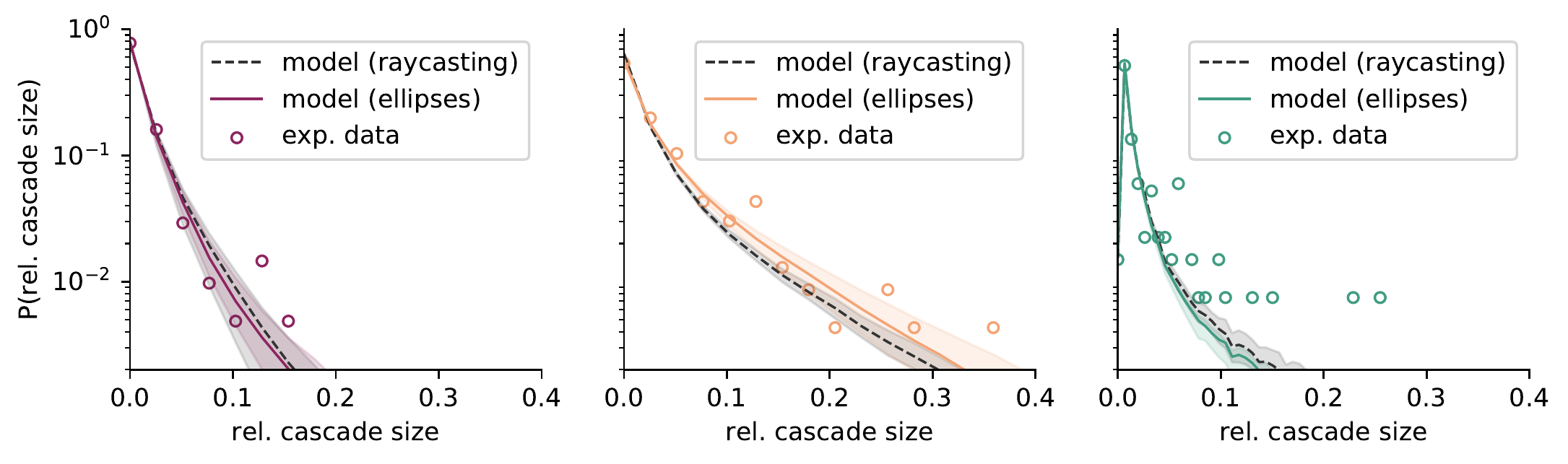}
        \caption{Best model fit (colored lines) for the different experimental dataset (dots) of cascade size distributions. The model is using networks based on ray casting (dashed, black lines) or the ellipse approximation (solid lines). Cascade size is measured relative to group size and shaded areas correspond to credible intervals for the model fit (i.e. the CI of the response threshold, see Fig. \ref{fig:ellipse_validation})}
        \label{fig:modelfitoneset}
    \end{figure}

\begin{figure}
        \centering
        \includegraphics[width=\linewidth]{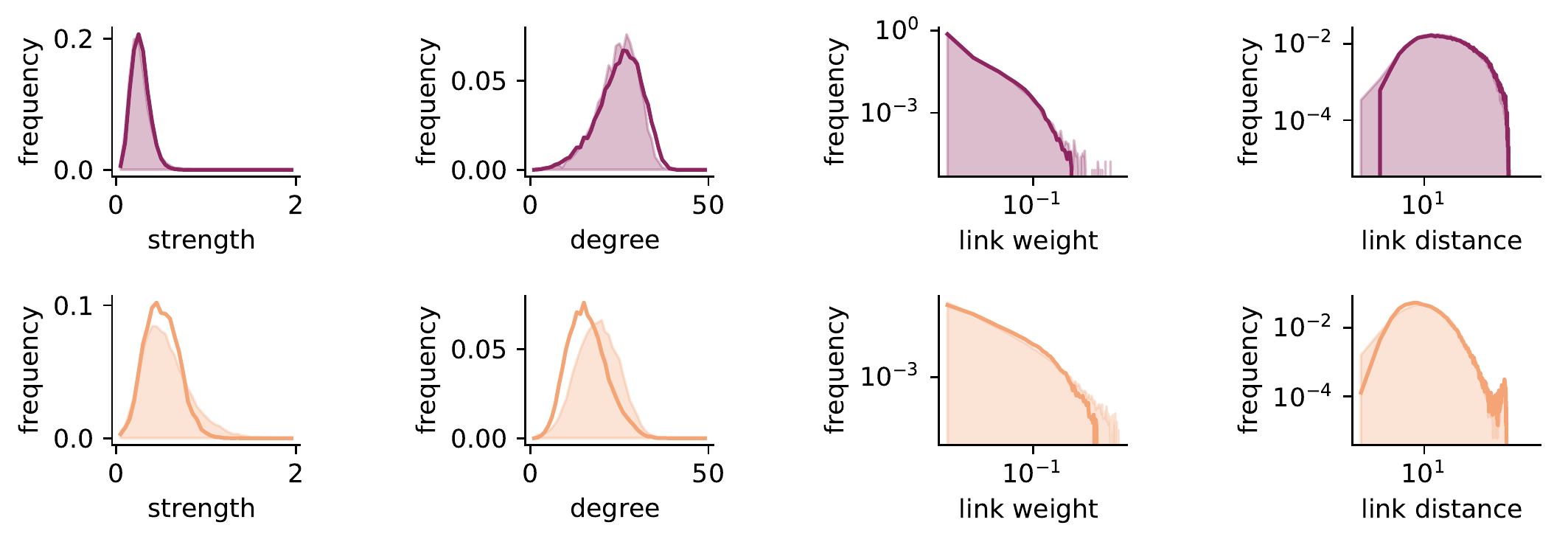}
        \caption{Comparison of networks constructed from original scale experimental data using the ellipse approximation with networks constructed from body pose estimation and ray casting (as in \cite{SosTwoBak19} and \cite{Rosenthal2015}) via distributions of different network properties. Filled area: Networks using ray casting. Line: Networks using ellipses. Histograms show (from left to right): strength (sum of all weighted network links of a node), degree (number of links per node), link weights, link distance (metric distance between two fish connected by a link). Each row corresponds to one of the data sets, top: 40 fish `Baseline', bottom: 40 fish `Alarmed'. Overall the approximation works well with only minor differences between the distributions. The elimination of overlaps in the ellipse networks also eliminates very small distances present in the school as seen in the right column of plots. This weakens some short and strong links by making them a bit longer (see link weight distribution). Additionally, the number of visible neighbors is decreased by this elimination of overlaps because fish that before were stacked on top of each other and not occluding each other's view now become close neighbors in the same plane and thus block a large part of each other's field of view. Especially in the case of the high density data set (40 fish `Alarmed') this difference is notable in the shift of the degree and strength distribution.}
        \label{fig:comparing_ellipse_network_props}

    \end{figure}

\begin{figure}
        \centering
        \includegraphics[scale=0.5]{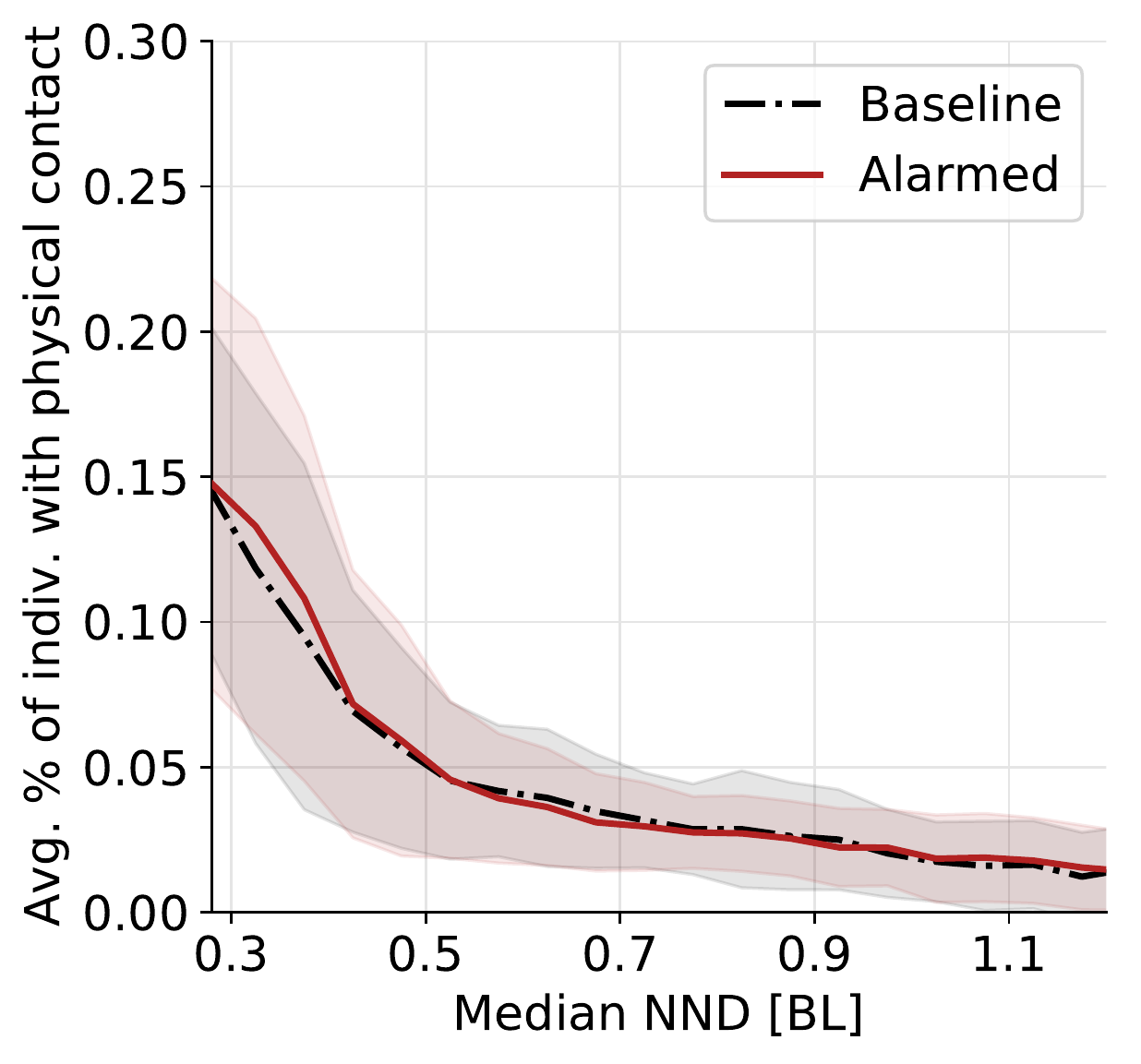}
        \caption{Investigating physical limit of density modulation: Estimation of average number of physical contacts an individual has with its neighbors at a certain density (given by the group's median nearest neighbor distance). Shaded areas indicate one standard deviation above and below the average. To estimate the number of contacts we identify ellipses that intersect when scaled up to 105\% of the original size.}
        \label{fig:touchpoints}
    \end{figure}

\begin{figure}
    \centering
    \includegraphics[scale=0.6]{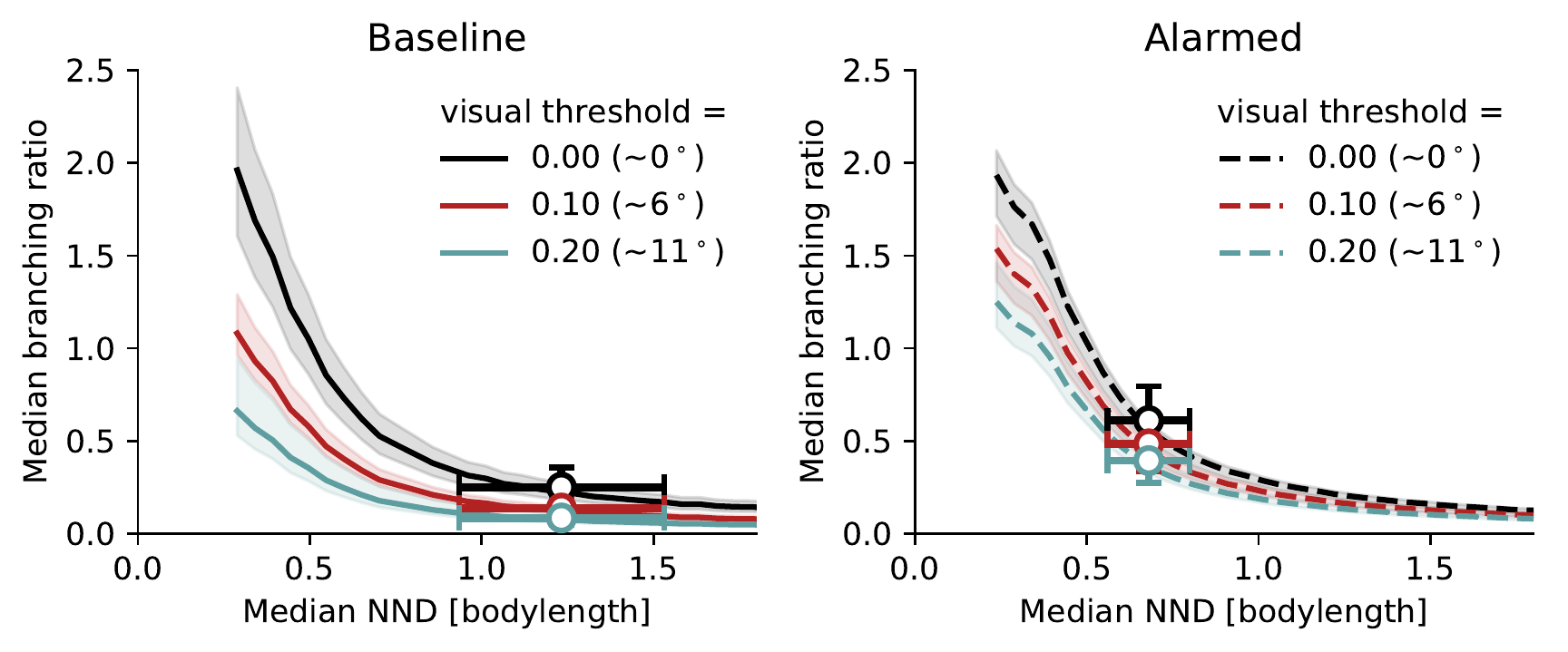}
    \caption{Branching ratio as function of median nearest neighbor distance (NND) for different visual tresholds (minimal angle of individual $i$ in the visual field of individual $j$ required for existence of network link between them). We observe a shift of the estimated critical point ($b=1$) to lower NND for increasing visual threshold, making it impossible to reach criticality without an additional change of individual responsiveness. This does not change our finding, that experimentally observed schools are subcritical, nor the general form of the relative individual payoff.}
    \label{fig:branching_visthresh}
\end{figure}

\begin{figure}
    \centering
    \includegraphics[scale=1]{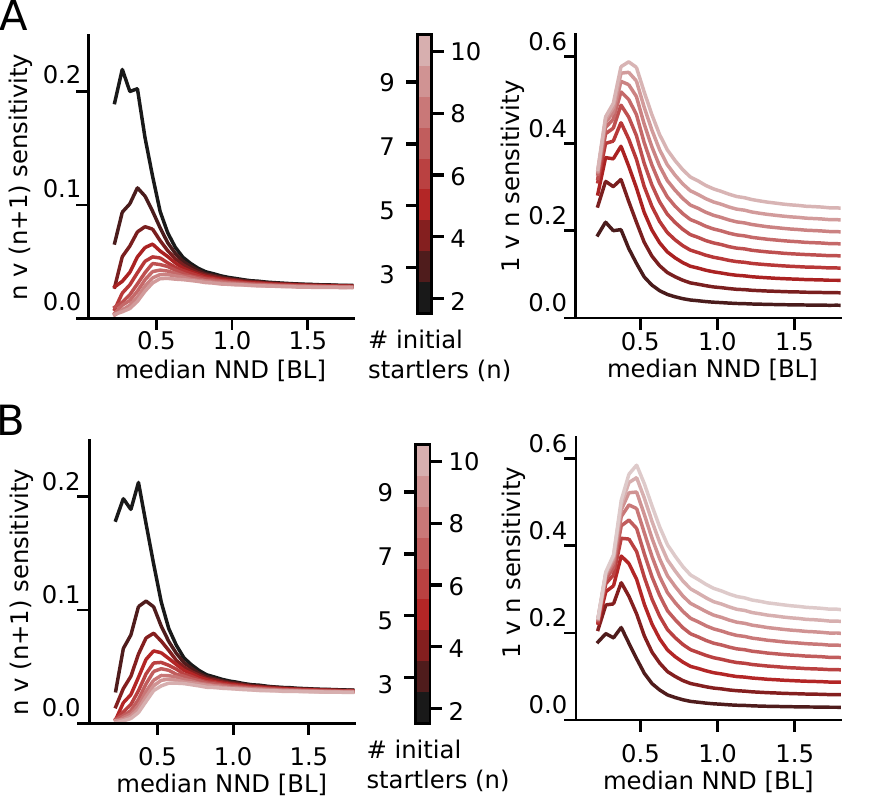}
    \caption{Alternative definitions of collective sensitivity as difference in average cascade size between $n$ and $n+1$ (left) or 1 and $n$ (right) initial startlers all exhibit a peak near criticality. A) Initial startlers are chosen randomly from the school. B) Initial startlers are randomly chosen network neighbors of the same (randomly chosen) individual. Results are qualitatively similar for both choices of initialization.}
    \label{fig:sensitivity_and_netneighbors}
\end{figure}

\begin{figure}
    \centering
    \includegraphics[width=0.55\linewidth]{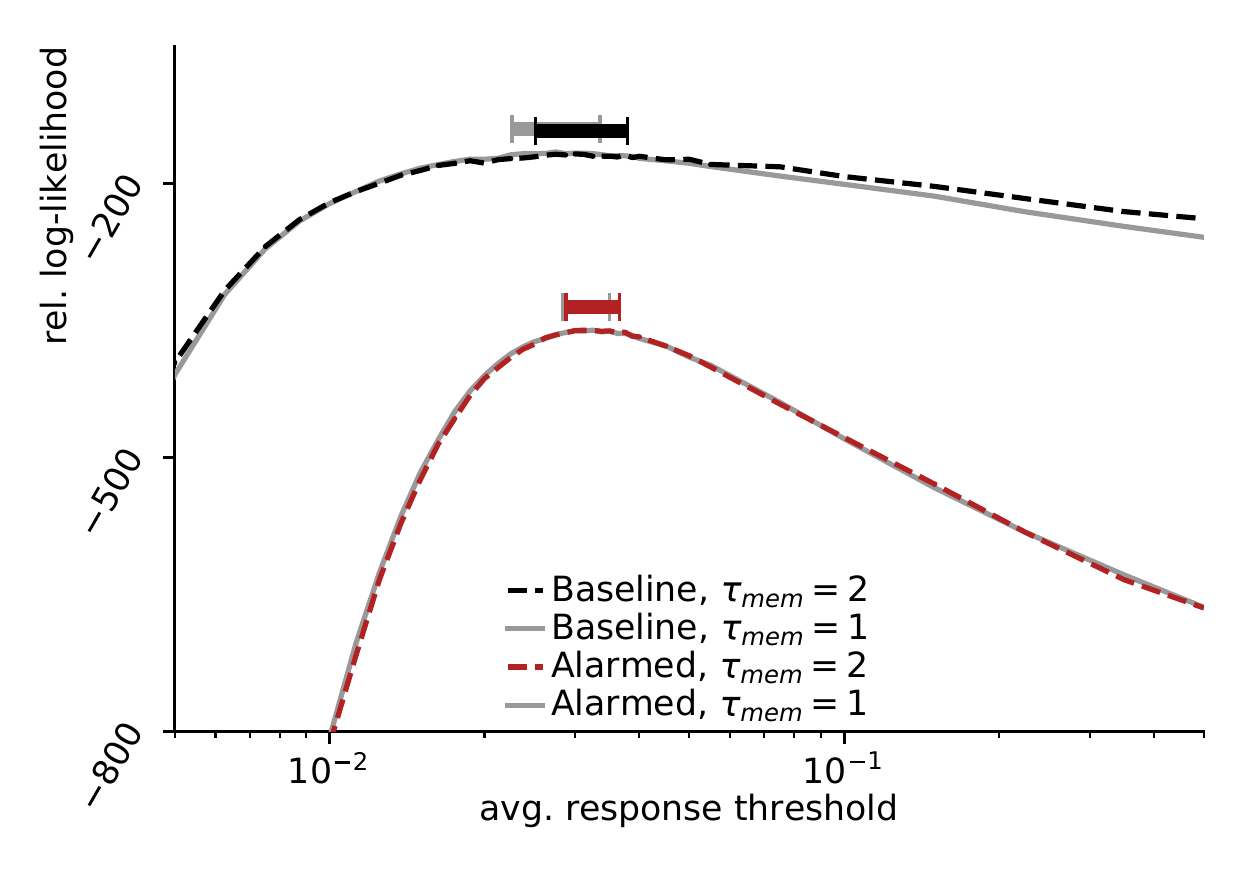}
    \caption{Relative log-likelihood for two different values of agent memory: $\tau_{mem}=1$ (dashed lines, as used in main text) and $\tau_{mem}=2$ (solid lines, as used in \cite{SosTwoBak19}). The optimal value of the avg. response threshold does not change significantly for the two choices.}
    \label{fig:agentmem}
\end{figure}

Figures \ref{fig:ellipse_validation} and \ref{fig:modelfitoneset} show the experimental data and the fitted model for the dataset of groups of $150\pm4$ fish from \cite{Rosenthal2015}. The coefficients of the logistic regression for the first responder probability are taken from the original study and given in table \ref{tab:betas}. Fig. \ref{fig:150fish_branching} shows the branching ratio for this dataset. The observed schools are subcritical.

The two datasets, while using the same species, differ in group size (40 vs. $150\pm4$ fish), experimental procedure and setup (i.e. handling of the fish prior to being placed in the tank, recording time, size- and age-matching, size of the tank and tank area per fish). This may explain the difference in average individual response threshold between the datasets.

\subsection{Discussion of model limitations and further experimental studies}
There are a number of limitations of the behavioral contagion model that could be addressed by more experimental data to improve and refine statistical analysis. Currently, interaction networks are built on pair-wise interactions fitted to first responder data. This assumes that interactions do not change during the cascade and that higher order interactions do not play a significant role. Additionally, with more experimental observations we could determine the features predictive of startle response and their relative importance for data confined to small density ranges and thus test if interaction rules depend on density. This would potentially show other types of interactions (acoustic, sensomotor) playing a role at higher densities. This detailed data for different (naturally occurring) densities could also provide a direct observation of the increase in average cascade size with increasing density. In an experimental setup that allows us to startle 1 or 2 initial startlers a detailed analysis of resulting cascade size distribution could potentially even show the increase in sensitivity with density as in figure 3B and C of the main text or Fig. \ref{fig:sensitivity_and_netneighbors}.
 
 The relative payoff measure is simplistic in the sense that it focuses solely on information processing which is only a single aspect of a wider range of important factors and lacks experimental evidence for the assumed reaction to a predator. Experiments with a real predator are needed to determine the number of initial startlers and potentially even see the density dependence of predator detection. Providing a stimulus in empty tank at different densities that occur via natural fluctuations may present another possibility test if density indeed has an effect on the detection ability. This lends itself to investigate another aspect: In our model we assume that the detection of the predator happens exclusively at the beginning of the cascade and the information then spreads only socially. In reality there will always be a direct detection of the predator happening along side the behavioral contagion process that reinforces the spreading and might be the reason that we observe such high false negatives in our model, which does not have this mechanism. In experiments with an artificial stimulus one could potentially change the duration for which a stimulus is shown to quantify this effect.

 \section{Sensitivity and branching ratio}
 \subsection{Alternative definitions of collective sensitivity}
  Fig. \ref{fig:sensitivity_and_netneighbors} depicts differences between average cascade sizes initiated by $n$ and $n+1$ or $1$ and $n$ initial startlers as a generalization of the definition in the main text. These sensitivities, while not directly motivated by the underlying theory of criticality, may be biologically relevant. We find, that independent of the used definition, we observe a maximum in collective sensitivity close to criticality. We note that the schools are most sensitive to differences for small numbers of initial startles. 
  \subsection{Branching ratio}
\begin{figure}
    \centering
    \includegraphics[width=0.8\linewidth]{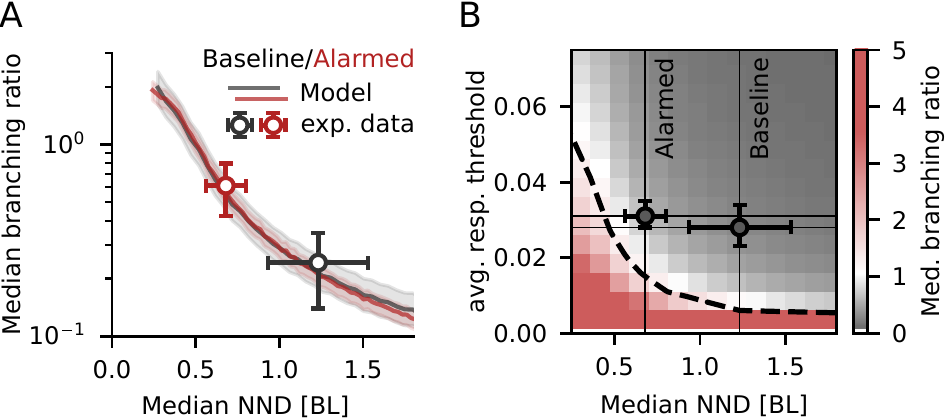}
    \caption{The average median branching ratio, an analytical estimate of criticality, as a function of A) median nearest neighbor distance (NND) and B) median NND and average response threshold. Lines in A) are averages over simulations for all rescaled networks binned by median NND. Shaded areas indicate the uncertainty of the model fit (average response threshold fit, see Table \ref{tab:LL_ellipse_fovea}). The dashed line in B) marks $b=1$ which is also include in Fig. 3C of the main text. Data points are averages over original scale networks and represent the experimentally observed schools. In A) both horizontal and  vertical error bars indicate one standard deviation of the average over networks in B) the vertical errorbars are the credible intervals for the model fit of the average response threshold, see table \ref{tab:LL_ellipse_fovea}.}
    \label{fig:branching}
\end{figure}

\begin{figure}
    \centering
    \includegraphics[width=0.8\linewidth]{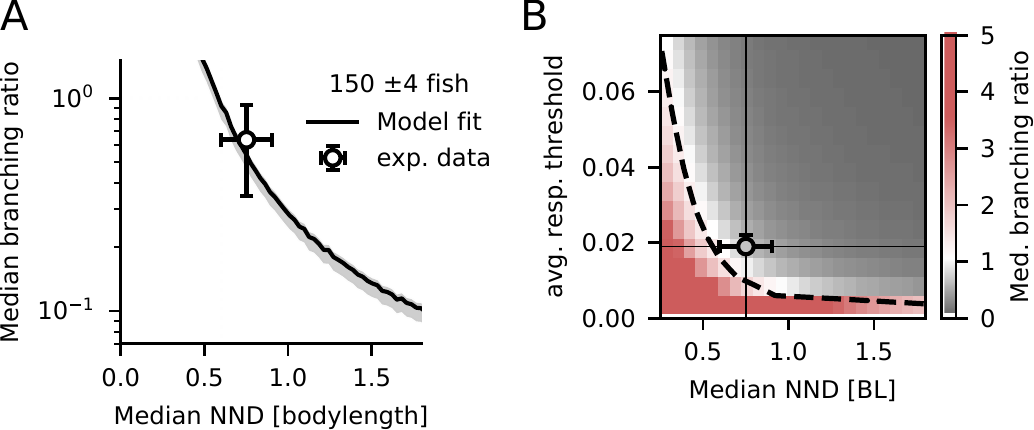}
    \caption{Branching ratio for dataset of approximately 150 fish from \cite{Rosenthal2015}. Model predictions (lines in A, colormap in B) and averages over original scale networks characterizing experimental observations (errobars, as in \ref{fig:branching}). The observed schools are subcritical, like the observed schools of 40 fish discussed in the main paper, with a slightly lower response threshold and an average median NND of 0.75$\pm$0.15\,BL.}
    \label{fig:150fish_branching}
\end{figure}

\begin{figure}
    \centering
    \includegraphics[scale=0.66]{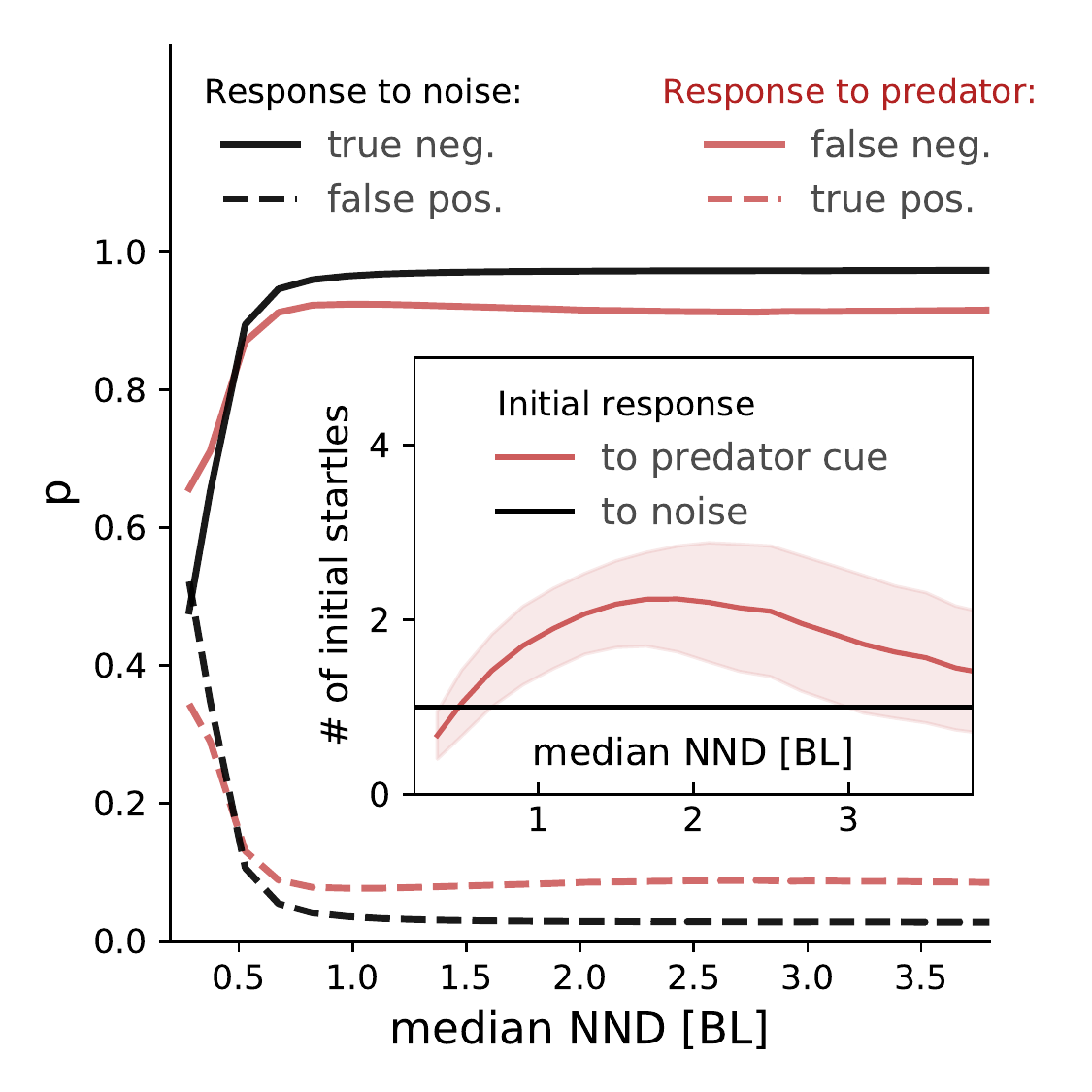}
    \caption{Likelihood of the different decision outcomes for the individual as a function of median nearest neighbor distance (NND). Close to the critical point false negatives decrease and false positives increase, and thus there remains a trade-off between two types of errors that can be managed according to the environment by choosing the appropriate distance to criticality. Inset shows the number of initial startlers used to trigger cascades assumed to be initiated by a predator cue (red) or noise (black).}
    \label{fig:errors}
\end{figure}

\begin{figure}
    \centering
    \includegraphics[width=0.9\linewidth]{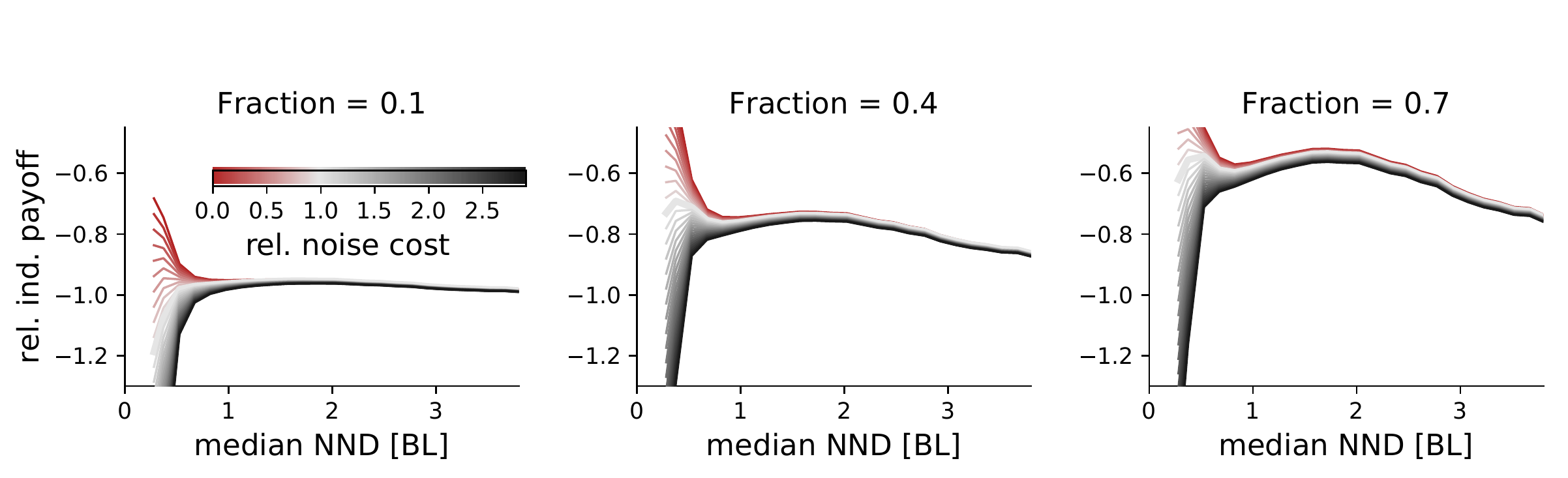}
    \caption{Relative payoff for different reactive fractions, $p_{detect}$, in the individuals that can see the predator. Increasing this fraction increases the maximum at intermediate densities but does not change the qualitative results. The left panel is as in the main paper. }
    \label{fig:SI_visualfraction}
\end{figure}

  The dependence of the branching ratio on median nearest neighbor distance and the average response threshold is shown in Fig. \ref{fig:branching} as well as the averages over original scale networks, characterizing the experimentally observed schools as subcritical.
  
 \section{Relative payoff measure}
\begin{figure}
    \centering
    \includegraphics[scale=1]{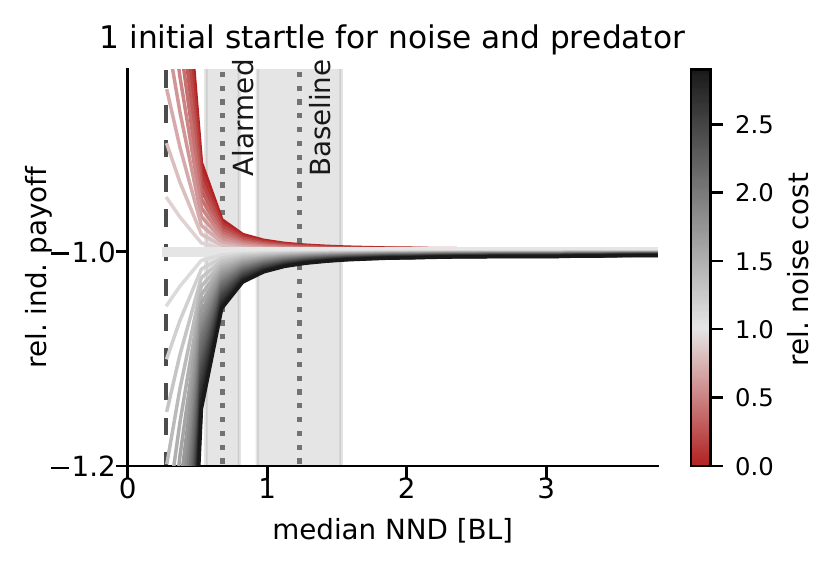}
    \caption{Relative payoff, assuming that both noise and predator cue trigger just one initial startle. The divergence at small NND remains, while the maximum at intermediate densities and the one at the critical point disappear.}
    \label{fig:fitness11}
\end{figure}

\begin{figure}
    \centering
    \includegraphics[width=0.35\linewidth]{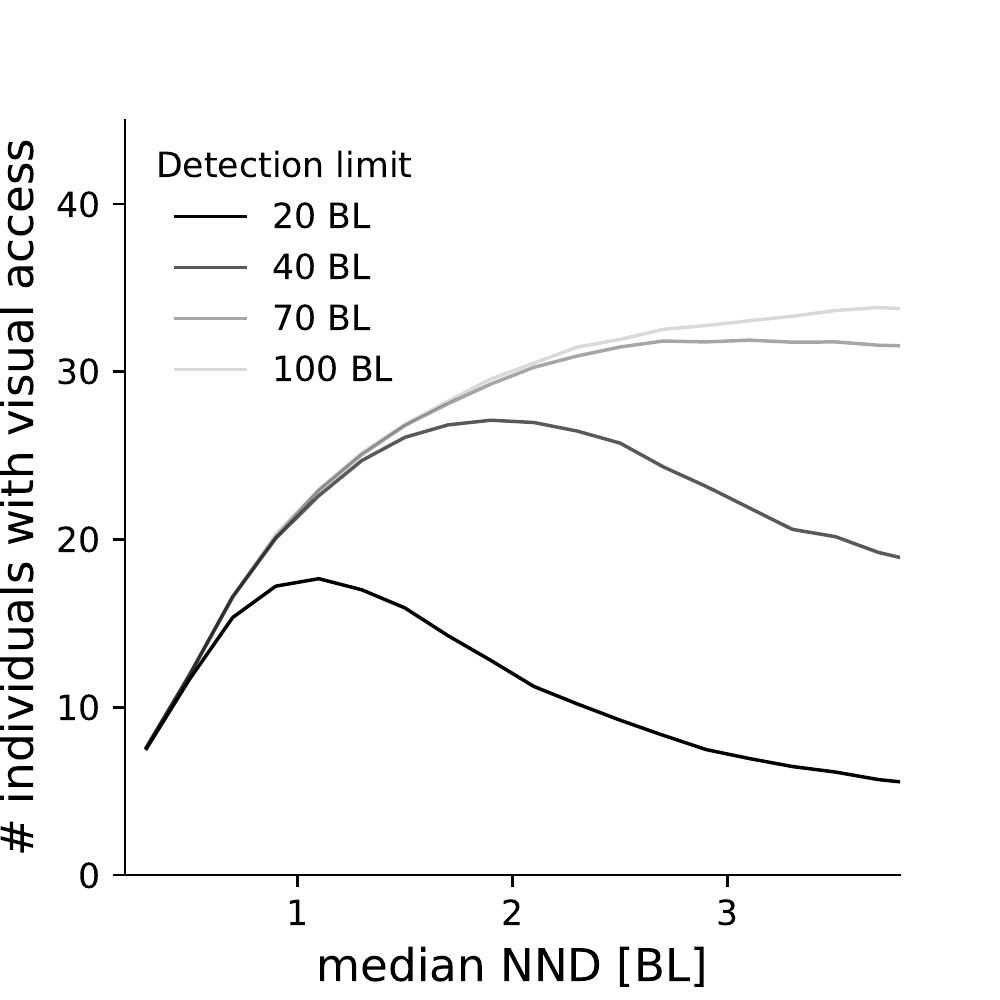}
    \includegraphics[width=0.35\linewidth]{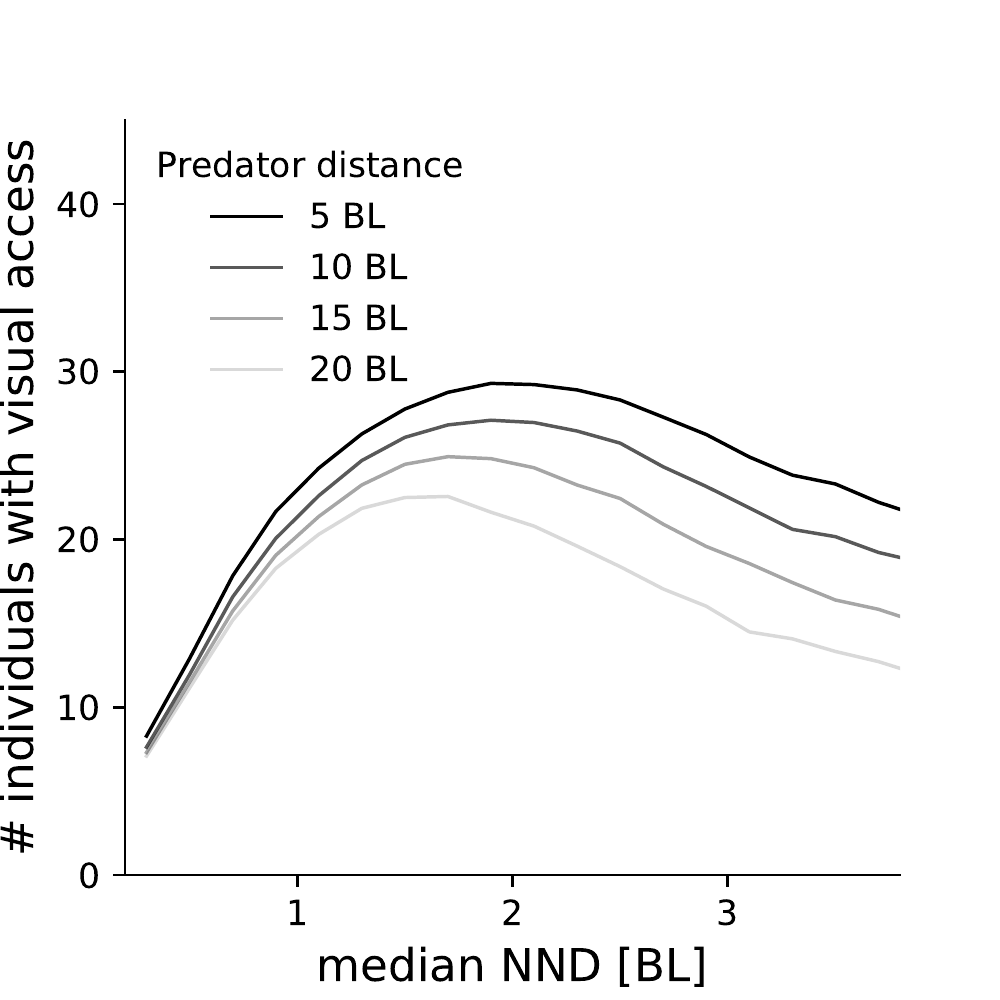}
    \caption{Influence of distance of group boundary to predator ($d_{pred}$, right) and maximal detection distance of the individual ($d_{max}$, left) on the number of individuals that have visual access to a predator plotted against median nearest neighbor distance. As long as there is an upper limit to the distance at which an individual can perceive a predator, the qualitative shape of the curve remains unchanged. The exact position of the maximum changes.}
    \label{fig:vis_detection}
\end{figure}

 We find in the main text that the relative payoff shows two local maxima. One is a result of the maximal visual detection of the predator by group members (as seen in Fig. \ref{fig:SI_visualfraction} where that maximum increases with increasing the fraction of individuals responding to their personal visual predator detection by startling). The other is due to the peak in sensitivity (as seen in the main text or Fig. \ref{fig:sensitivity_and_netneighbors}) to the number of initial startlers at the critical point (see Fig. \ref{fig:fitness11}, where the initial response to predator and noise cue is assumed to be identical and there is no maximum at criticality). The position of the maximum payoff depends on the relative noise cost but also on the parameters of the visual predator detection as we will explain here before giving a more detailed derivation of the payoff measure.
 
 \subsection{Parameters of visual predator detection}
\label{sec:vis_detection}
\begin{figure}
    \centering
    \includegraphics[width=.9\linewidth]{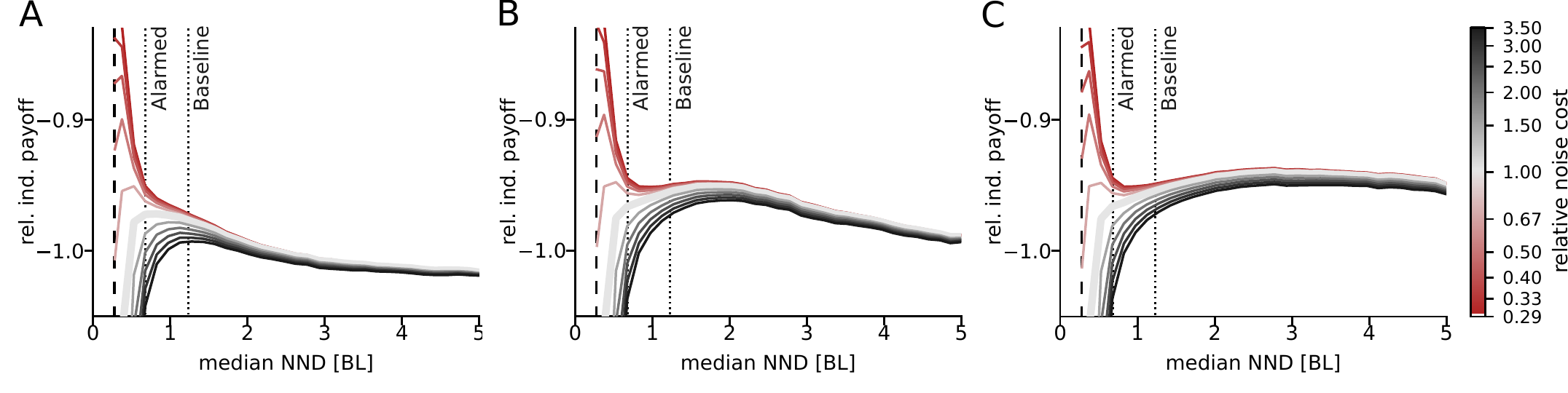}
    \caption{Relative payoff for varying parameter $d_{max}$ (maximum detection limit) of the visual detection for $d_{max}=20$ (left), $d_{max}=40$ (middle, as in main text) and $d_{max}=70$ (right). All plots use $d_{pred}=10$. Due to the shift in the position of the maximum of visual access (see Fig. \ref{fig:vis_detection}, left plot) the position of the second maximum of the relative payoff also shifts. For $d_{max}=20$ (left) both maxima (the criticality-based one and the visual-access-based one) merge into a single maximum.}
    \label{fig:SI_payoff_params}
\end{figure}

\begin{figure}
    \centering
    \includegraphics[width=0.45\linewidth]{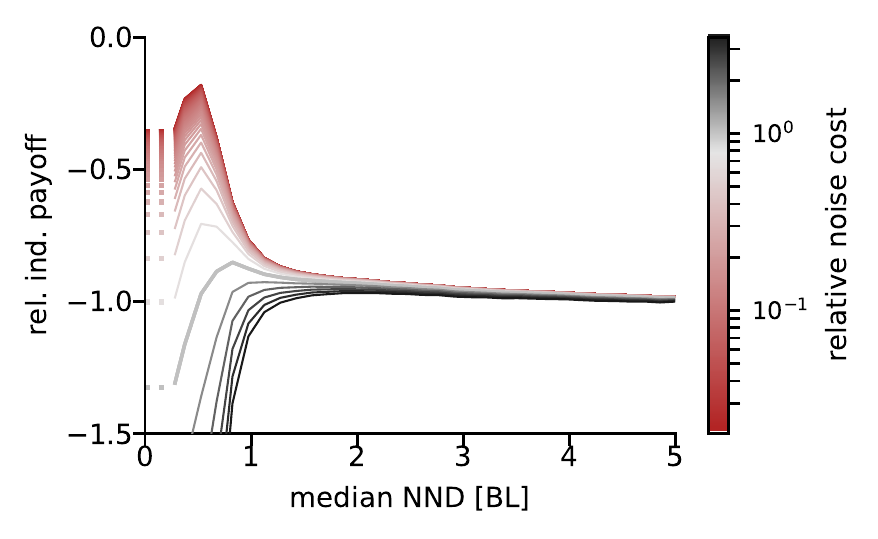}
    \includegraphics[width=0.45\linewidth]{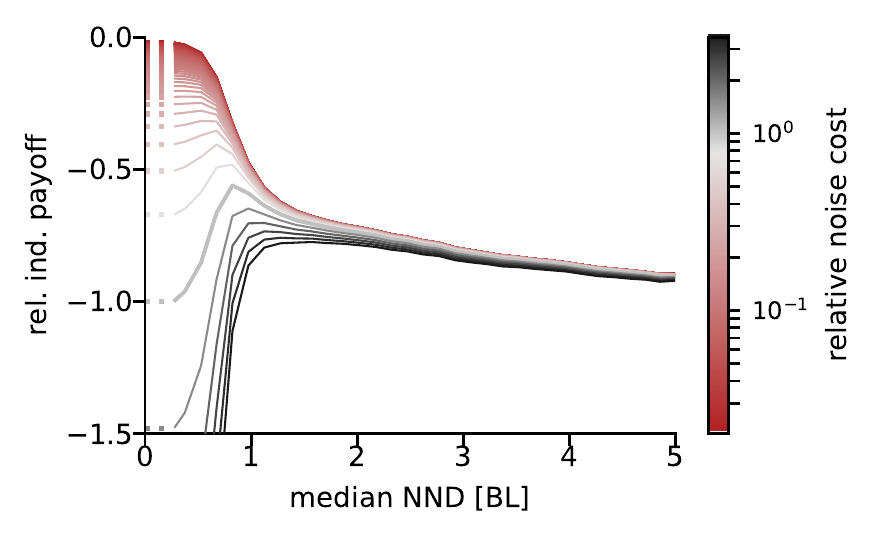}
    \includegraphics[width=0.45\linewidth]{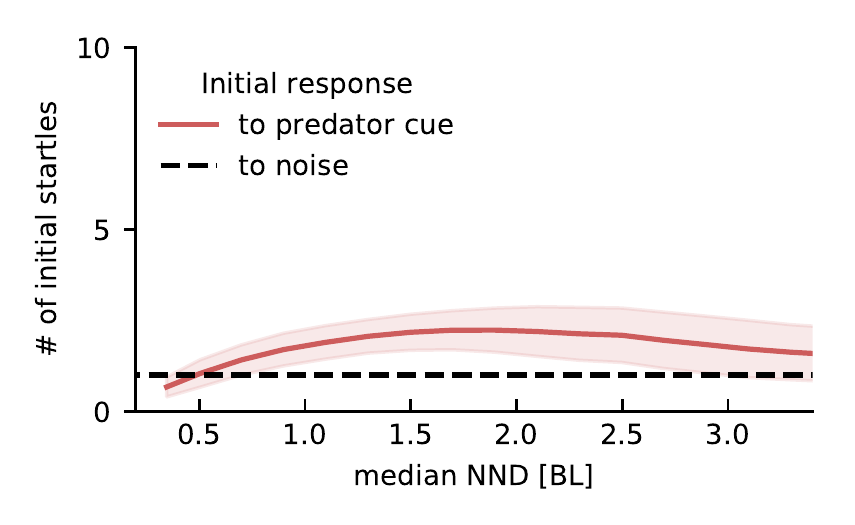}
    \includegraphics[width=0.45\linewidth]{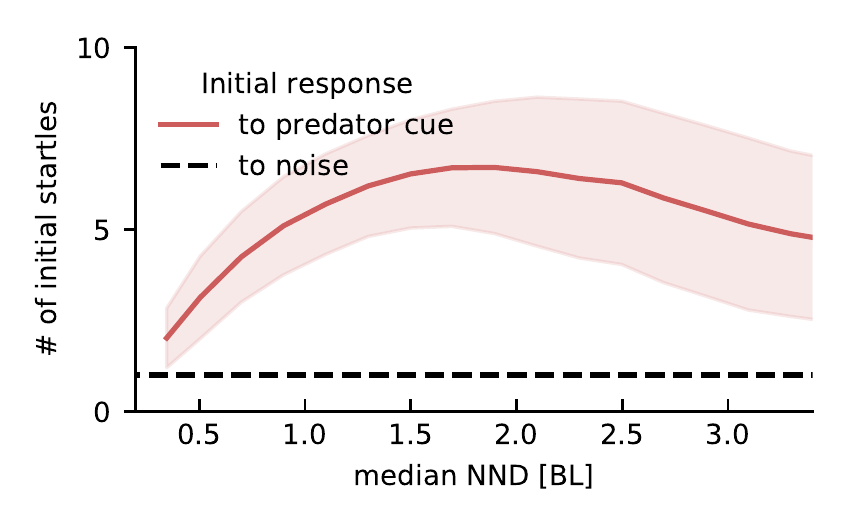}
    \caption{Payoff for an average response threshold of $\bar{\theta}=0.009$ and reactive fraction $a=0.1$ (left) or $a=0.3$ (right). Because at low NND the number of initial startlers goes below 1 for $a=0.1$, the supercritical state (very low NND), is not optimal. Even though every initial startle yields a global response in this regime, the likelihood of a predator being detected is lower than one and thus no cascade may be initiated at all. This decrease in payoff for very low NND and very low relative noise cost (red curves) disappears as soon as we assume that at least one individual will respond to a predator (see right plots). The existence of a maximum at criticality for intermediate relative noise cost is unchanged by this. Unlike for $\bar\theta\approx0.03$ as used in the main text, here the second maximum disappears because the critical point has shifted to higher NND and thus closer to the maximum of the visual detection (bottom plots).}
    \label{fig:SI_payoff_lowthresh}
\end{figure}

\begin{figure}
     \centering
     \includegraphics[width=\linewidth]{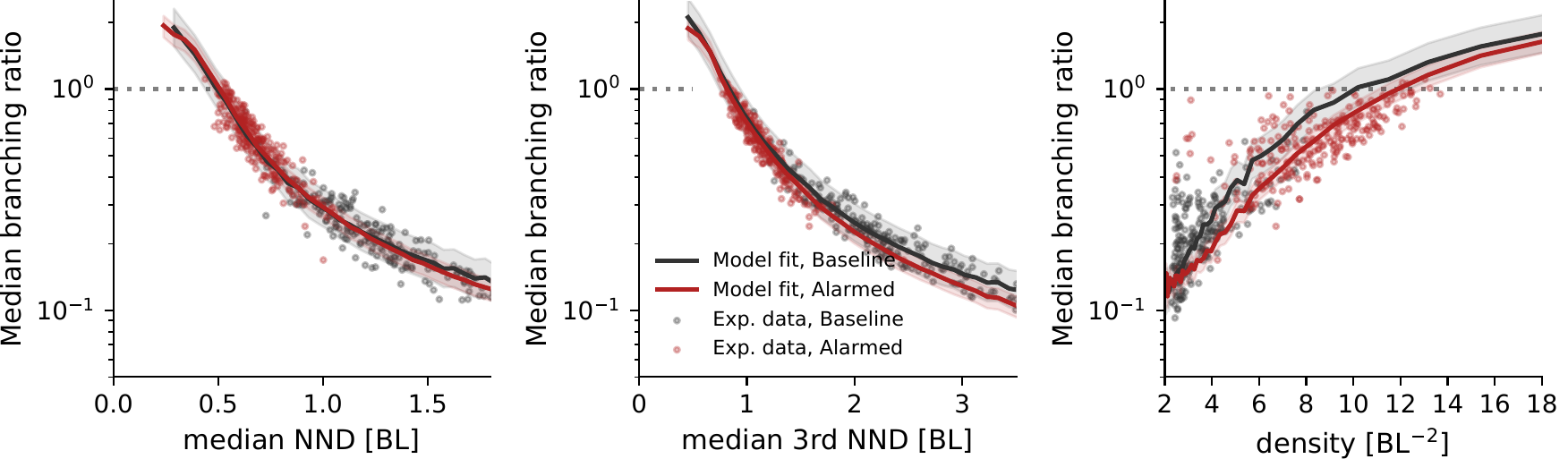}
     \caption{Average median branching ratio plotted against three measures of density: median nearest neighbor distance (left, as used in main text), third nearest neighbor distance (middle) and density (right, calculated as ratio of group size and area of the group's convex hull). Data points represent results using the networks obtained from experimental observations without rescaling of inter-individual distances. Independent of the density measure the observed schools are on average subcritical.}
     \label{fig:densitymeasures}
 \end{figure}

Our model of visual predator detection depends on three parameters, namely the distance of the predator from the group boundary, $d_{pred}$, the maximal distance at which a predator is still visible $d_{max}$ and the response probability of an individual $p_{detect}$ (probability to startle given that the individual has visual access to the predator). Figures \ref{fig:SI_visualfraction} and \ref{fig:SI_payoff_lowthresh} show the influence of $p_{detect}$ on the relative payoff. Fig. \ref{fig:vis_detection} illustrates how $d_{max}$ and $d_{pred}$ influence the number of individuals that can see the predator, Fig. \ref{fig:SI_payoff_params} how they influence the relative payoff. In Fig. \ref{fig:SI_payoff_params}A one does not observed curves with two separate maxima but just one maximum shifting from a density optimizing individual access to visual information of the predator to the critical density, optimizing sensitivity to number of initial startlers. The observed median NNDs of the `Baseline' and the `Alarmed' dataset are optimal for a relative noise cost of 3. and 1. respectively.
 \subsection{Construction of the relative payoff measure}
 \label{sec:SI_payoff_derivation}
 Here, we start by considering all possible behavior-environment combinations (false and true positives and false and true negatives, marked $fp$, $tp$, $fn$ and $tn$ respectively), their costs $c$ and the rates with which they occur $\rho$. All of them are added up into a payoff rate defined as
\begin{align}
   \tilde{\psi} = \rho_{fp}~c_{fp}+\rho_{fn}~c_{fn}+\rho_{tp}~c_{tp}+\rho_{tn}~c_{tn}
    \label{eq:general_payoff}
\end{align}
We assume that a predator causes $N_{init}=n(\text{NND})$ initial startlers while a noise cue causes just one, $N_{init}=1$, and cues indicating predators appear at a rate $\rho_p$ while cues indicating no predator appear at rate $\rho_n$. Then, each individual has the following rates of
\begin{align}
\begin{aligned}
    \text{false pos.:}\quad \rho_{fp}&=\rho_n~p(S|N_{init}=1),\\
    \text{false neg.:}\quad \rho_{fn}&=\rho_p~\Big[1-p\big(S|N_{init}=n(\text{NND})\big)\Big],\\
    \text{true pos.:}\quad \rho_{tp}&=\rho_p-\rho_{fn},\\
    \text{true neg.:}\quad \rho_{tn}&=\rho_n-\rho_{fp},\\
\end{aligned}
\label{eq:errorprobs}
\end{align}
where $p(S|N_{init})$ means the probability to startle (as part of the cascade) given $N_{init}$ initial startlers and can be obtained from simulations. Inserting \eqref{eq:errorprobs} into \eqref{eq:general_payoff} yields
\begin{align}
    \begin{aligned}
    \tilde{\psi}=& 
    \rho_pc_{tp}+\rho_nc_{tn}+\\&\rho_{fp}(c_{fp}-c_{tn})+\rho_{fn}(c_{fn}-c_{tp}).
    \end{aligned}
\end{align}
Since we are interested in the relative payoff of different school densities for a fixed environment and do not want to compare payoff rates between different environments, we can choose the baseline freely. We set it to $\rho_pc_{tp}+\rho_nc_{tn}$, the rate at which an individual on average gains benefits by making correct decisions. By rescaling the payoff rate in units of $\rho_p(c_{fn}-c_{tp})$ (the average payoff rate associated to a predator cue) we can define a relative payoff as
\begin{align}
\begin{aligned}
    \psi&=\frac{\tilde{\psi}-(\rho_pc_{tp}+\rho_nc_{tn})}{\rho_p(c_{fn}-c_{tp})}\\[0.4cm]
    &=\frac{\rho_{fp}}{\rho_p}\left(\frac{c_{fp}-c_{tn}}{c_{fn}-c_{tp}}\right)+\frac{\rho_{fn}}{\rho_p}
    \end{aligned}
\end{align}
Inserting \ref{eq:errorprobs} then yields
\begin{equation}
    \psi=r~p(S|N_i=1)+[1-p(S|N_i=n(\text{NND})]
\end{equation}
with
\begin{equation}
r=\frac{\rho_n}{\rho_p}\left(\frac{c_{fp}-c_{tn}}{c_{fn}-c_{tp}}\right)
\end{equation}
Here, $\rho_n/\rho_p$ is the relative prevalence of  noise cues compared to the predator cues. If $\rho_n/\rho_p\gg1$ the environment is very noisy, if $\rho_n/\rho_p\ll1$ there is a lot of predation. In the cost-based term $\frac{c_{fp}-c_{tn}}{c_{fn}-c_{tp}}$ the numerator measures the costs associated to startling behavior, the denominator quantifies the costs of an attack that are due to injury or risk of death. This term can be thought of as the relative costs of noise. The combination of both terms, $r$, can best be described as the \textit{relative noise cost}.


    \begin{table}
         \centering
         \renewcommand{\arraystretch}{1.2}
         \begin{tabular}{c|c|c|c|c}
             dataset & visual field method & avg. response thresh. & CI & max. LL \\\hline
             40 fish baseline & ray casting & 0.021 & [0.018,0.033] & -175\\
             40 fish baseline & ellipses & 0.028 & [0.023,0.034] & -166\\
             40 fish `Alarmed' & ray casting & 0.032 &[0.030 0.034] & -387\\
             40 fish `Alarmed' & ellipses & 0.031 &[0.028,0.035] & -360\\
             150 fish & ray casting & 0.016 & [0.014,0.018] & -305\\
             150 fish & ellipses & 0.019 &[0.018,0.022] & -297\\
         \end{tabular}
         \caption{Fitting the response threshold for networks constructed from the different methods of visual field construction. Credible intervals overlap for both methods and maximum likelihood is comparable. Fitting is based on 10000 simulations per network with the initial startler set in accordance with the experimentally observed cascade initiator.}
         \label{tab:LL_ellipse_fovea}
     \end{table}
 \begin{table}
        \centering
        \begin{tabular}{c|c|c|c}
             data set &$\beta_1$ (Intercept)& $\beta_2$ (LMD coefficient)& $\beta_3$ (RAA coefficient)\\\hline
             40 fish joined & -0.271 & -2.737 &-0.097\\
             150 fish &0.302 &-3.272 (-1.421) &-0.126 
        \end{tabular}
        \caption{Coefficient values to equation \eqref{eq:pij} as determined by the logistic regression of first response rates. The base of the logarithm is 10 as in \cite{SosTwoBak19}. The values for 150 fish were taken from \cite{Rosenthal2015} which was using a natural log (coefficient value in bracket) and was transferred to log base 10.}
        \label{tab:betas}
    \end{table}

    \begin{table}
    \renewcommand{\arraystretch}{1.2}
     \centering
    \begin{tabular}{r c c c c c}
    & abs. vis & avg.&\\
     dataset & thresh.  &  response thresh. & CI & max. LL \\ \hline
     40 fish `Baseline' & 0.00 & 0.027& [0.022, 0.033] & -164\\  
      & 0.02 & 0.028& [0.023, 0.034] & -166\\
     & 0.10  &0.049&[0.041, 0.055] & -159\\ 
     & 0.20  &0.080&[0.056, 0.100]&-173\\ \hline
    40 fish `Alarmed' & 0.00 & 0.031& [0.029, 0.035] & -360\\  
      & 0.02 & 0.031 & [0.028, 0.035]& -360\\ 
      & 0.10 & 0.039& [0.036, 0.044]&-355\\ 
      & 0.20 & 0.048& [0.041, 0.054] &-367\\
    \end{tabular}

    \caption{Influence of a visual threshold (minimal angle required for visibility) on fitting the average response threshold based on 10000 simulation runs for each network. These values of the average response threshold are used to calculate the branching ratio, see fig. \ref{fig:branching_visthresh} and equation \eqref{eq:branching_SI}.}
    \label{tab:visthresh}

    \end{table}
    





\end{document}